\pdfoutput=1
\documentclass[12pt,twoside]{amsart}
\usepackage{amssymb}
\usepackage{graphicx}
\usepackage[english]{babel}
\usepackage[utf8]{inputenc}
\usepackage{natbib}
\usepackage{tikz}
\usepackage{caption}
\usepackage{subcaption}
\usetikzlibrary{calc}
\usepackage{enumitem}
\usepackage{fancyhdr}

\def\RR{\mathbb{R}}    
\def\EE{\mathbb{E}}    
\def\R{\mathcal{R}}   

\def\proof{\noindent{\em Proof.}~}
\def\eproof{\mbox{\ }\hfill$\square$}

\newtheorem{theorem}{Theorem}
\newtheorem{lemma}{Lemma}
\newtheorem{proposition}{Proposition}
\newtheorem{corollary}{Corollary}
\newtheorem{assumption}{Assumption}
\newtheorem{definition}{Definition}
\newtheorem{remark}{Remark}

\usepackage{a4wide}

\title[Welfare at Risk]{Welfare at Risk: Distributionally Robust Bounds for Policy Evaluation}

\author{Costas Lambros \and Emerson Melo }
\thanks{Department of Economics, Indiana University Bloomington; email:\{clambros,emelo\}@iu.edu}
\date{\today}
\begin{document}
	
	\begin{abstract}This paper develops a framework to study how the welfare effects of policy interventions are distributed across individuals when those effects are not directly observed. We bound the superquantile of individual welfare changes by the superquantile of conditional average welfare changes, which we can typically identify, revealing how gains and losses spread across the population. We also bound the share of the population whose welfare loss from the policy exceeds any given threshold --- a distribution-free counterpart to average cost--benefit analysis, closer to a poverty-rate measure of policy harm than to a mean welfare effect. We further extend both bounds to allow for ambiguity in the analyst's knowledge of the distribution of observable characteristics: a non-robust bound estimated on one population can be systematically overoptimistic when applied to a different target population, while the robust version, with its ambiguity radius chosen commensurate with the shift, restores a valid guarantee. We illustrate the method in generalized Roy models with subjective participation costs, bounding not only the distribution of welfare changes in the overall population but also, in a result new to this literature, the entire distribution of welfare (and of participation costs) among those who select into treatment --- the treatment-on-the-treated distribution, not merely its mean. We connect these results to the multi-outcome IV framework of \cite{Heckman_Urzua_Vytlacil_2006_REStat}, showing that our robust bounds remain valid across all instruments whose implied covariate distributions lie within the ambiguity set, which addresses the instrument-sensitivity problem they document.\end{abstract}
	\maketitle
	\vspace{3ex}

	\thispagestyle{empty}
	
	\newcommand{\spacing}[1]{\renewcommand{\baselinestretch}{#1}\large\normalsize}
	\textwidth      5.95in \textheight 600pt
	\spacing{1.5}

	\section{Introduction}\label{Sec:Intro}
	
	Samuelson-Bergson welfare analysis of policy interventions typically focuses on average gains or losses, often overlooking their distributional consequences (e.g., \cite{bhattacharya2024nonparametric}). Mean welfare changes can mask substantial heterogeneity across individuals: a policymaker may want to know which subpopulations are most affected by an intervention, or whether it treats them fairly. Addressing this requires methods that capture distributional variation, such as quantile-based welfare analysis, heterogeneous-agent frameworks, or inequality measures like the Atkinson index and Gini coefficient.
	\smallskip
	
	Assessing distributional heterogeneity requires knowing the distribution of individual welfare, which is typically unidentified for two reasons. First, the analyst often observes only aggregate market data, so only average welfare changes are identified (\cite{Berry1994,BerryLevinsohnPakes1995,Berry_Haile_ECMA2014}, \cite{Allen_Rehbeck}). Second, even when individual data are available, each agent is observed in only one of two potential-outcome states---treated or untreated---so the individual welfare change, and hence its distribution, cannot be recovered (\cite{HECKMAN2007_Handbook_partI}).
	\smallskip
	
	In this paper, we develop a distributional welfare framework that accounts for the heterogeneity in welfare generated by policy interventions. In particular, we derive quantile-based bounds that allow the analyst to learn about the distribution of individual welfare changes, even when this distribution is unobserved or unidentified.
	\smallskip
	
	Our approach builds on the perturbed utility model (PUM) of \cite{Allen_Rehbeck} and \cite{McFadden_Fosgerau_2012}, a latent choice model that accommodates general additive unobserved heterogeneity and nests several well-known frameworks, including the additive random utility model (ARUM) developed in \cite{mcfadden1972conditional, McFadden1978, mcf1}.\footnote{Beyond ARUM, the  PUM class also encompasses bundling models (\cite{Gentzkow_AER_2007}; \cite{iaria2020identification}; \cite{fox2017note}), consumer choice models with latent constraints (\cite{agarwal2022demand}), and demand systems with additive unobserved heterogeneity (\cite{brown2003strong}; \cite{Matzkin_2007}).} ARUM is widely used for its tractability and its foundations in discrete choice theory, which facilitate counterfactual analysis (\cite{Berry1994}; \cite{BerryLevinsohnPakes1995}; \cite{Berry_Haile_ECMA2014}), and it plays a central role in the treatment-effects literature, where self-selection is modeled as a discrete choice problem. This structure underpins the marginal treatment effect (MTE), which makes explicit the roles of observables and unobservables in individuals' decisions to participate in a social program (\cite{Bjorklund_Moffitt1987}; \cite{Heckman_Vytlacil_ECMA_2005_MTE, HECKMAN2007_Handbook_partI}; \cite{Heckman_Urzua_Vytlacil_2006_REStat, Heckman_Urzua_Vytlacil}).
	\smallskip
	
	We restrict our analysis to environments in which the analyst observes only \emph{aggregate market data} or estimates of conditional average welfare changes (such as average treatment effects) are identified. In such settings, the distribution of individual welfare changes is not identifiable, motivating the need for a framework that delivers informative bounds. As shown in \cite{Allen_Rehbeck}, the PUM framework identifies conditional average treatment effects, a key requirement for our results.
	\smallskip
	
	An important concern in our distributional welfare analysis is related to its robustness \citep{Hansen_Sargent_2001, Hansen_Sargent2008}: the welfare bounds above depend on the distribution of observable covariates, which is rarely known with certainty. The target population may differ from the study population, the covariate distribution must itself be estimated, and, in instrumental-variable settings, the relevant population weights depend on the choice of instrument \citep{Heckman_Urzua_Vytlacil_2006_REStat,Heckman_Urzua_Vytlacil}. We address this by extending our framework to allow for distributional (model) ambiguity.  In particular, we bound welfare changes under the \ emph {worst-case} distribution over a $\phi$-divergence neighborhood where the center is given by the nominal covariate distribution (usually the empirical distribution). 
	
	\smallskip
	
	From a technical standpoint, our framework relies on the concept of superquantiles, developed in \cite{Rockafellar2000OptimizationOC, ROCKAFELLAR20021443} and \cite{Rockafellar_Royset}. Superquantiles, also known as conditional value at risk (CVAR),  generalize quantiles by averaging over the distribution's tail beyond a given quantile---for instance, the 20\% superquantile reports the average welfare among the bottom 20\%---giving a richer view of distributional impacts than a quantile alone. For our robustness analysis, we draw on the literature on distributionally robust optimization \citep{Kuhn_Shafiee_Wiesemann_2025}. Formally, we use convex duality to represent the robust problem, an infinite-dimensional optimization program, as a finite-dimensional concave optimization problem that is tractable and solvable with conventional optimization packages.

	\subsection{Contributions} 
	We make several contributions. First, Theorem \ref{Bound_Change_Utility} shows that the superquantile of individual welfare changes---unobservable to the analyst---is bounded above by the superquantile of conditional average welfare changes, which is identified. Under the appropriate relabeling, this inequality is mathematically equivalent to Theorem 3.1 in \cite{Kallus2023} (established there for individual and conditional-average treatment effects under \emph{binary} treatment and unconfoundedness), and likewise for our Theorem \ref{Bound_Change_Utility_Lower} and his Theorem 3.3; we do not claim novelty for these two inequalities themselves, and return to this point, along with the genuine differences between the two papers, in Section \ref{sec:lit}. Our contribution is to show that such bounds apply to the general PUM class, and to draw out what they deliver: information on conditional average welfare changes (or average treatment effects, in the Roy model) is enough to characterize, for any $\beta$-quantile, the average gain or loss among the $(100\times \beta)\%$ worst affected by a policy, identify the subpopulations most adversely affected, and assess the efficacy and fairness of alternative interventions. Both bounds are computable via linear programming. We further show that the same inequality, holding uniformly across $\beta$, implies second-order stochastic dominance of conditional average over individual welfare changes (Corollary \ref{Cor:SSD}). More importantly, we show that the same basic inequality can be \emph{inverted} to obtain a bound on the buffered probability of exceedance \citep{Rockafellar_Royset_2010,Mafusalov_Uryasev_2018} (Proposition \ref{Prop:bPOE}): rather than fixing a tail fraction $\beta$ in advance, an analyst can instead fix a policy-relevant loss threshold---a benefit cliff, a poverty line, a compensation trigger---and bound the \emph{fraction} of the population that falls below it, answering how many people are hurt this badly directly from identified, aggregate-level data.
	
	\smallskip
	
	Our second contribution extends Theorems \ref{Bound_Change_Utility} and \ref{Bound_Change_Utility_Lower} to allow for ambiguity in the analyst's knowledge of the distribution of observables $X$. Theorems \ref{Robust_Bound_Change_Utility} and \ref{Robust_Bound_Change_Utility_Lower} in Section \ref{Sec:Robust_Bounds} show both bounds remain valid when the nominal distribution $F$ of $X$ is replaced by a worst-case over a $\phi$-divergence neighborhood of $F$, and Proposition \ref{Robust_CVAR_Dual_Prop} gives a tractable dual that reduces computing the robust bound to a finite-dimensional concave program. Section \ref{Sec:Sim_Layer2} verifies this numerically: a non-robust bound estimated on one population can be systematically overoptimistic once the policy is applied to a different target population, for instance after a distributional shift, and the robust bound restores a valid guarantee once $\rho$ (the statistical distance from the nominal distribution) is chosen commensurate with the shift.
	\smallskip

	Our third contribution applies the framework to an economic environment where distributional welfare is central to policy analysis: social program evaluation in generalized Roy models with subjective participation costs (\cite{Heckman_Vytlacil_ECMA_2005_MTE}), extending the cost--benefit framework of \cite{Eisenhauer_Heckman_Vytlacil_2015} to incorporate explicit distributional considerations. Although the distributions of benefits, costs, and \emph{individual} welfare are not directly observable, we show our bounds to be informative about these quantities, leveraging the identification results in \cite{Eisenhauer_Heckman_Vytlacil_2015} to recover informative features of these unobserved distributions at the individual level.
	\smallskip
	
	Specifically,  we derive bounds for the distributions of costs, benefits, and welfare among agents who participate in the program --- \emph{treatment on the treated} (TT). To our knowledge, these results, which condition on realized treatment status ($D=1$) rather than on a policy assignment, are novel within generalized Roy models with participation costs. We give this result its own dedicated treatment in the body of the paper.
	\smallskip
	
	In addition, we connect our results to the multi-outcome IV framework of \cite{Heckman_Urzua_Vytlacil_2006_REStat}. A central finding in Section~VI is that IV estimands in ordered and unordered choice models are weighted averages of transition-specific MTEs, with weights that can be negative, so point-identified IV-based welfare estimates are sensitive to instrument choice. Our distributionally robust bounds (Theorems~\ref{Robust_Bound_Change_Utility} and \ref{Robust_Bound_Change_Utility_Lower}) address this sensitivity: because the infimum is taken over an ambiguity set rather than over instrument choices, the robust welfare risk is a valid bound for every instrument whose implied covariate distribution lies within divergence $\rho$ of the nominal one, and the radius this requires can be computed from estimated propensity scores. 
	
	\smallskip
	
	The rest of the paper unfolds as follows. Section \ref{sec:lit} discusses the related literature. Section \ref{Sec:PUM_Section} briefly presents the relevant aspects of the PUM framework and formalizes superquantiles, discussing their key properties. Section \ref{Sec:Bounds_Ind_Welfare} develops upper and lower bounds that let us learn about the distribution of individual-level welfare changes, which are often unidentified, and Section \ref{Sec:Robust_Bounds} extends these bounds to account for ambiguity in the distribution of $X$. Section \ref{Sec:Simulations} presents simulations that provide evidence for, and demonstrate the usability of, the results in Section \ref{Sec:Bounds_Ind_Welfare}. Section \ref{Sec:Applicability_Distributional_WDZ} examines the applications, leading with program evaluation with subjective participation costs and treatment on the treated. Section \ref{Sec:HUV_Connection} connects our framework to the multi-outcome IV results of \cite{Heckman_Urzua_Vytlacil_2006_REStat}, showing that our DRO bounds address the instrument-sensitivity problem documented there. Section \ref{Sec:Conclusion} concludes.
	
	\section{Related literature}\label{sec:lit} 
	Welfare analysis in latent utility models has received significant attention: \cite{mcf1} introduced the general ARUM framework and established the existence of a representative agent; \cite{Small_Rosen_1981} adapted traditional welfare economics to discrete choice models; \cite{Hanneman_1996} examined income effects on ARUM welfare changes; \cite{Mcfadden_1996} developed computational techniques for mean compensated variation; and \cite{McFadden_Fosgerau_2012} generalized demand and welfare analysis to the PUM framework. All of these studies, however, focus on average welfare measures, overlooking distributional aspects.
	
	\smallskip
	
	Much work has been done on non-parametric identification of latent utility models. \cite{Allen_Rehbeck} provides non-parametric identification of average welfare changes within the PUM framework used here, requiring only aggregate data---a requirement our results share---but, like the traditional welfare-analysis literature, remains confined to \textit{average} welfare measures. Our results are complementary: we build on their non-parametric identification to go beyond average analysis and characterize the distribution of welfare effects at the individual level.
	\smallskip
	
	\cite{Bhattacharya_2015} studies welfare in non-additive random utility models (NARUMs) and identifies the marginal distributions of compensated variation (CV) and equivalent variation (EV) using conditional choice probabilities. Both their analysis and ours provide non-parametric tools for distributional welfare evaluation, but differ in two main respects: their results are specific to NARUMs, whereas our PUM framework allows for richer demand patterns, including substitutability and complementarity; and \cite{Bhattacharya_2015} requires individual-level choice data, whereas---as with \cite{Allen_Rehbeck} above---our approach accommodates the standard setting of aggregate data alone (\cite{Berry_Haile_ECMA2014}). Our framework also extends well beyond demand analysis,  as discussed in the application in Section \ref{Sec:Applicability_Distributional_WDZ}.

	\smallskip 
	
	The closest paper to ours is \cite{Kallus2023}. Like us, he uses superquantiles to recover features of the individual welfare distribution from conditional average treatment effects. At the level of the basic inequalities, the two papers overlap directly: our Theorem \ref{Bound_Change_Utility} is, under the appropriate relabeling, the same inequality as Theorem 3.1 in \cite{Kallus2023}, and our Theorem \ref{Bound_Change_Utility_Lower} is likewise equivalent to his Theorem 3.3; both are proved by the same iterated-expectations-and-Jensen argument. We make this equivalence explicit rather than claim novelty, since neither inequality depends on any feature of the PUM, the welfare interpretation, or unobserved heterogeneity in choice --- as we note explicitly following the proof of Theorem \ref{Bound_Change_Utility}, the argument applies to any integrable random variable and its conditional mean. The contributions of our paper relative to \cite{Kallus2023}---none of them addressed there or elsewhere in the treatment-effect-risk literature---lie in four respects. First, we work with the broader class of PUMs, covering a wide range of demand and behavioral models, and---again exploiting the aggregate-data identification noted above---show that the resulting superquantile bounds can be operationalized without individual-level data. Second, in Section \ref{Sec:Robust_Bounds} we extend both bounds to allow for $\phi$-divergence ambiguity in the distribution of $X$ (Theorems \ref{Robust_Bound_Change_Utility} and \ref{Robust_Bound_Change_Utility_Lower}). \cite{Xu_Yang_2025} also bring distributional robustness to treatment-effect analysis, but differently: they place a Wasserstein ball on the joint distribution of the potential outcomes and derive a minimax mean-squared-error predictor of the individual treatment effect, without covariates. Our ambiguity set is instead a $\phi$-divergence ball on the distribution of $X$, and our target is a superquantile of individual welfare changes rather than a mean-squared-error criterion.  Third, we develop a distributional analysis for generalized Roy models with subjective treatment costs, including the distribution of welfare among actual participants; these last two applications lie outside \cite{Kallus2023}'s scope. Four, even at the level of the shared basic inequality, we show in Proposition \ref{Prop:bPOE} that it can be inverted into a bound on the buffered probability of exceedance \citep{Rockafellar_Royset_2010,Mafusalov_Uryasev_2018}---the fraction of the population at risk of a policy-relevant loss threshold, rather than the magnitude of loss for a fixed tail fraction; \cite{Kallus2023} states his bounds only in CVaR-magnitude form, with no such inversion.
	
	\section{A Robust Distributional Framework}\label{Sec:PUM_Section}
	This section has two objectives. First, we introduce the perturbed utility model (PUM) of \citet{Allen_Rehbeck} and \citet{McFadden_Fosgerau_2012}. We use the PUM primarily as a welfare foundation: its utility-maximization structure provides a natural economic interpretation of the individual welfare change,
	$\tau(X,t,\varepsilon)$, where $X$ denotes observed covariates, $t$ the intervention, and $\varepsilon$ unobserved heterogeneity, together with its conditional expectation $\tau(X,t)$. These objects constitute the fundamental welfare primitives throughout the paper. The PUM encompasses a broad class of latent utility models with additively separable unobserved heterogeneity, including additive random utility models (ARUMs), bundling models, and matching models. Our main application, the generalized Roy model, is a special case. Because the PUM is closely connected to standard discrete choice theory, it provides an economically meaningful interpretation of the welfare bounds developed below.
	
	Second, we develop a distributional framework for welfare analysis in PUMs that goes beyond average treatment effects. To this end, we briefly review quantiles and introduce superquantiles, which play a central role in our analysis of distributional welfare.
	\smallskip
	
	\subsection{The PUM approach}Consider a decision-maker (DM) making a utility-maximizing choice among an unordered set of alternatives $j\in\mathcal{J}=\left\{1,\ldots, J\right\}$.  The DM's optimal choice vector  $Y$ satisfies:
	
	\begin{equation}\label{PUM_Model}
		Y\in\arg\max_{y\in B}\sum\limits_{j=1}^Jy_ju_j(X_j)+\R(y,\varepsilon)
	\end{equation}
	where $B\subseteq\RR^J$ denotes the DM's budget constraint and $X_j=(X_{j,1},\dots,X_{j,d_j})'$ denotes the observed characteristics of good $j$, which has $d_j$ observed characteristics. These good-specific regressors are collected into $X=(X_1',\dots,X_J')'$, and $X\subset \mathcal{X}$. We note that there can be common regressors across $X_j$ and $X_k$, $j\neq k$ and $j,k\in\mathcal{J}$. Therefore, $X$ can encode observed characteristics of both the goods \textit{and} the DM. The vector $\vec{u}=(u_1,\dots,u_J)$ encodes how the desirability of a good varies with its regressors. Moreover, $\R$ is the \textit{regularizer function} and depends on the DM's choice, $y$, and the agent's unobserved heterogeneity, $\varepsilon\in E$.
	\smallskip
	
	To better see how  PUM  serves as an umbrella for a large class of models, consider a straightforward ARUM.  In an  ARUM environment, a DM must choose one of the choices $j\in\mathcal{J}$, whose utility is given by
	$$v_j=u_j(X_j)+\varepsilon_j$$
	Setting $B=\Delta_J$,  where $\Delta_J$ is the $J$-dimensional simplex defined as:
	$$\Delta_J\triangleq \Big\{y\in\RR^J|\sum\limits_{j=1}^Jy_j=1,\quad y_j\geqslant0\forall j\in\mathcal{J}\Big\}$$
	and regularizer function $\R(y,\varepsilon)=\sum\limits_{j=1}^Jy_j\varepsilon_j,$
	it follows that the ARUM framework corresponds to a particular instance of a PUM. Furthermore, in the ARUM case  it is well-known that under the assumption that $\varepsilon$ is absolutely continuous, the DM will choose one of the available options $j\in \mathcal{J}$, which corresponds to choosing one of the vertices of $\Delta_J$.\smallskip
	
	While the ARUM is the leading special case, the scope of formulation (\ref{PUM_Model}) extends well beyond discrete choice. First, the choice vector $Y$ need not be interpreted as a vector of choice probabilities: when $B$ is a general closed convex set, $Y$ may represent \textit{quantities} of goods, so that the PUM describes average demand systems with additively separable heterogeneity (\cite{Allen_Rehbeck}). Second, the same perturbed structure arises in matching markets with transferable utility, where the equilibrium matching maximizes a social surplus function of PUM form (\cite{Galichon_Salanie_2021}), and in models of stochastic choice, for which \cite{Fudenberg_et_al_2015} provide an axiomatic foundation. The structure also accommodates moral hazard environments in which an agent selects a distribution over outcomes subject to a convex effort cost (\cite{Hebert2018}). Because our distributional results below rely only on the structure in (\ref{PUM_Model}) and Assumption \ref{Assumption_PUM1}, they apply to each of these environments.\smallskip

	In general, we will focus on the case where $B=\Delta_J$ as above. However, most of our results apply to the general case of $B$ being a closed convex subset of $\RR^J$. Following \cite{Allen_Rehbeck}, we use the following assumption. 
	\begin{assumption}\label{Assumption_PUM1}
		Assume the following:
		\begin{enumerate}[label=(\roman*)]
			\item The random variables $Y$, $X$ and $\varepsilon$ satisfy (\ref{PUM_Model}).
			
			\item $B\subseteq\RR^J$ is a nonempty,  closed set. 
			\item $u_j:\RR^{d_j}\rightarrow\RR$ for $j\in\mathcal{J}$ where $d_j$  denotes the dimension of the space of covariates associated to choice $j\in \mathcal{J}$.
			\item $\R:\RR^J\times E\rightarrow \RR\cup\{-\infty\}$, where $E$ denotes the support of $\varepsilon$, is an extended real-valued function.
		\end{enumerate}
	\end{assumption}
	

	\smallskip
	
	Throughout the  paper, we  make use of the following notation:
	$$U(y;x,\varepsilon)\triangleq\sum_{j=1}^Jy_ju_j(x_j)+\R(y,\varepsilon)$$ 
	for all $y\in B$ and $X=x$. When $X$ is treated as a random variable, we use $U(y;X,\varepsilon)$.
	\smallskip
	
	\cite{Allen_Rehbeck} show that the  PUM aggregates. Based on their aggregation result, we can define the conditional average welfare $W(x)$ as:
	$$W(x)\triangleq \EE\left[\max_{y\in B}U(y;x,\varepsilon)\mid X=x\right].$$ 
	
	The previous expression generalizes the social surplus function introduced by \cite{mcf1} in the context of ARUMs. More importantly for our purposes, \cite{Allen_Rehbeck} show that welfare changes, measured by $W(x^1)-W(x^0)$, are nonparametrically identified using only aggregate market data. We exploit this result as a key building block of our analysis.
	\subsection{Superquantiles} Instead of focusing on average welfare (or average welfare changes) across the entire distribution of $\varepsilon$, our focus will be on average welfare among a specific subset of the population, namely the $(100 \times \beta) \%$-worst affected.  To formalize this notion, we use quantiles and the lesser-known concept of superquantiles introduced in \cite{Rockafellar2000OptimizationOC,ROCKAFELLAR20021443} and \cite{Rockafellar_Royset}.
	\smallskip
	
	Let $Z$ be a random variable with finite mean. Its $\beta$-quantile is
	\[
	F_Z^{-1}(\beta)=\inf\{\lambda:F_Z(\lambda)\geq\beta\},
	\]
	where $F_Z(z)=\mathbb{P}(Z\leq z)$. The $\beta$-superquantile
	\citep{Rockafellar2000OptimizationOC}$,$ corresponding to the average of the
	$(100\times\beta)\%$ lowest outcomes, is defined as
	\begin{equation}\label{Lower_CVAR_Equation_Def}
		\mathbb{S}^F_{\beta}(Z)
		\triangleq
		\sup_{\lambda\in\RR}
		\left\{
		\lambda+\frac{1}{\beta}\EE_F[Z-\lambda]_{-}
		\right\},
	\end{equation}
	where $[t]_{-}\triangleq\min\{t,0\}$. The supremum is attained at any
	$\beta$-quantile of $Z$.
	More generally, when $Z$ may have point masses,
	\citet{ROCKAFELLAR20021443} show that
	\begin{equation}\label{Optimal_value_S}
		\mathbb{S}^F_\beta(Z)
		\in
		\left[
		\EE_F\!\left[Z\,\big|\,Z<F_Z^{-1}(\beta)\right],
		\,
		\EE_F\!\left[Z\,\big|\,Z\leq F_Z^{-1}(\beta)\right]
		\right].
	\end{equation}
	
	Furthermore, if  $Z$ is a continuous random variable, then
	\[
	\mathbb{S}^F_\beta(Z)
	=
	\EE_F\!\left[Z\,\big|\,Z\leq F_Z^{-1}(\beta)\right].
	\]

	The superquantile is continuous in $\beta$, concave, monotone,
	translation invariant, and positively homogeneous
	\citep{Shapiro_et_al_2013}. These properties make it the correct way to
	generalize the average of the $(100\times\beta)\%$ lowest outcomes
	to discontinuous distributions, as also emphasized by \cite{Kallus2023}
	in the treatment-effect setting. More importantly, superquantiles
	provide a tractable way to incorporate distributional heterogeneity
	into welfare evaluation and, as we show below, into risk-averse
	treatment allocation.

	\begin{figure}
		\centering
		\begin{tikzpicture}
			\draw[->] (-4.5,0) -- (4,0) node[right] {$z$};
			\draw[->] (0,-0.1) -- (0,4) node[above] {$f_Z(z)$};
			
			\draw[thick, black, domain=-4.3:0, samples=100, smooth]
			plot (\x, {3.4*exp(-0.5*(\x/1.55)*(\x/1.55))});
			\draw[thick, black, domain=0:3.6, samples=100, smooth]
			plot (\x, {3.4*exp(-0.5*(\x/0.95)*(\x/0.95))});
			
			\begin{scope}
				\clip (-4.5,0) rectangle (-1.8,4);
				\fill[blue!25] (-4.3,0)
				plot[domain=-4.3:-1.8, samples=100, smooth] (\x,{3.4*exp(-0.5*(\x/1.55)*(\x/1.55))})
				-- (-1.8,0) -- cycle;
			\end{scope}
			\node[font=\tiny] at (-2.65,0.18) {worst $20\%$};
			
			\draw[dashed] (-1.8,0) -- (-1.8,1.73);
			\node[below] at (-1.8,-0.08) {\footnotesize $F_Z^{-1}(\beta)$};
			
			\draw[dashed, red, thick] (-3.0,0) -- (-3.0,1.05);
			\node[below, red] at (-3.0,-0.65) {\footnotesize $\mathbb{S}_\beta(Z)$};
		\end{tikzpicture}
		\caption{The $\beta$-superquantile versus the $\beta$-quantile ($\beta=0.2$ shown). The quantile $F_Z^{-1}(\beta)$ is only the cutoff below which the worst $(100\times\beta)\%$ of outcomes fall; the superquantile $\mathbb{S}_\beta(Z)$ is the \emph{average} outcome within that tail, and therefore lies weakly below it ($\mathbb{S}_\beta(Z)\leq F_Z^{-1}(\beta)$). For a left-skewed distribution of welfare changes, as illustrated here, the gap between the two can be substantial, since $\mathbb{S}_\beta(Z)$ reflects the severity of losses in the tail, not merely their frequency.}
		\label{Figure_Superquantile}
	\end{figure}
	\smallskip 
	
	For intuition, Figure \ref{Figure_Superquantile} illustrates the superquantile of a left-skewed random variable $Z$, representing a distribution of welfare changes with a heavy lower tail. Note that $f_Z(\cdot)$ denotes the probability density function of $Z$ while $F_Z(\cdot) $ denotes the cumulative distribution function of $Z$. For a given $\beta$, the quantile $F_Z^{-1}(\beta)$ marks the cutoff below which the worst $(100\times\beta)\%$ of outcomes fall, while the lower superquantile $\mathbb{S}^F_\beta(Z)$ is the average outcome within that shaded tail, and therefore lies weakly below the quantile. While this section has focused on lower superquantiles due to their relevance to the rest of the paper, it is straightforward to adapt \eqref{Lower_CVAR_Equation_Def} to capture the right tail of the distribution associated with $Z$. The adaptation comes from the fact that the right superquantile can be defined as  $\bar{\mathbb{S}}^F_{\alpha}(Z)=-\mathbb{S}^F_{1-\beta}(-Z),$ where $\alpha=1-\beta$. In particular,  $\bar{\mathbb{S}}_{\alpha}(Z)$ can be written as:
	$$\overline{\mathbb{S}}^F_\alpha(Z)=\min_{\lambda\in \mathbb{R}}\left
	\{\lambda+{1\over 1-\alpha}\mathbb{E}_F[Z-\lambda]_+\right\}$$
	where $(t)_{+}=\max\{t,0\}$.

	\subsection{The buffered probability of exceedance } 
	
	Rather than bounding the \emph{magnitude} of welfare loss for a fixed tail fraction $\beta$, an analyst may instead want to bound the \emph{fraction of the population at risk} of a specific, policy-relevant loss threshold $\theta$---for instance, a welfare loss exceeding a given dollar amount. This is precisely the buffered probability of exceedance of \citet{Rockafellar_Royset_2010}, adapted here to our lower-tail convention.
	
	\begin{definition}[Buffered probability of exceedance]\label{Def:bPOE}
		For a random variable $Z$ and a threshold $\theta\in(\operatorname{ess\,inf}Z,\,\EE[Z]]$, the buffered probability that $Z$ falls short of $\theta$ is
		\begin{equation}\label{bPOE_def}
			p_\theta(Z)\;\triangleq\;\inf\{\beta\in(0,1]:\mathbb{S}_\beta(Z)\geq \theta\}.
		\end{equation}
	\end{definition}
	
	Since $\beta\mapsto\mathbb{S}_\beta(Z)$ is continuous and nondecreasing \citep{Shapiro_et_al_2013}, $p_\theta(Z)$ is well defined on this domain and identifies the smallest tail fraction whose average outcome reaches $\theta$; following \citet{Mafusalov_Uryasev_2018},  $p_\theta(Z)$ is interpreted as the natural inverse of the superquantile mapping, in the same sense that a CDF inverts a quantile function. As we shall see, in the context of distributional welfare analysis, $p_\theta(Z)$ provides valuable information about the size of the population for which welfare change falls below the threshold $\theta$.
	
	\smallskip
	
	The definitions above implicitly assume the analyst knows the distribution $F$. The following subsection relaxes this assumption, introducing \emph{robust superquantiles} that remain valid under ambiguity about $F$.
	
	\subsection{A Distributionally Robust Framework}\label{Sec:DRO_Framework} We consider here an environment where the analyst does not know the true distribution of the observable covariates $X$. Throughout, the object of interest is a random variable $Z$ that is a function of $X$---specifically, $Z=\tau(X,t)$ or, holding the conditional law of $\varepsilon\mid X$ fixed at its nominal value, $Z=\tau(X,t,\varepsilon)$---so uncertainty about the distribution of $Z$ arises entirely from uncertainty about $F_X$, and we state the robustification below in terms of $F_X$.  The nominal distribution $F\equiv F_X$ should be understood as an estimate or a best guess of the true, unknown distribution of $X$; the conditional law of $\varepsilon$ given $X$ is left unrestricted by the robustification, since the identification results of Section \ref{Sec:Bounds_Ind_Welfare} below condition on $X$ throughout. Recognizing this ambiguity, we analyze the following problem:
	\begin{equation}\label{Robust_CVAR}
		\mathbb{S}^R_\beta(Z)=\inf_{G\in\mathcal{U}(F,\rho)}\mathbb{S}^G_\beta(Z) 
	\end{equation}
	where $\mathcal{U}(F,\rho)$ represents the \emph{uncertainty set} of distributions for $X$ that are at \emph{statistical distance} $\rho>0$ from the \emph{nominal} distribution $F$. 
	\smallskip
	
	Intuitively, problem (\ref{Robust_CVAR}) models a situation where nature selects the worst-case distribution $G$ for $X$ from the set $\mathcal{U}(F,\rho)$. From a mathematical standpoint, the set $\mathcal{U}(F,\rho)$ can be quite complex and potentially infinite-dimensional (for instance, when $X$ is continuous). To overcome this technical hurdle, the literature on distributionally robust optimization (\cite{Shapiro}, \cite{Shapiro_et_al_2013}, \cite{Kuhn_Shafiee_Wiesemann_2025}) has focused on different ways of modeling $\mathcal{U}(F,\rho)$ using different notions of  \emph{statistical distances}.
	\smallskip

	In this paper, we measure the distance between two probability distributions for $X$ using the so-called $\phi$-divergence (\cite{Csiszar1967}, \cite{LieseVajda1987}, and \cite{Pardo2005}). Let $\phi: \mathbb{R} \rightarrow(-\infty,+\infty]$ be a proper closed convex function such that $\mbox{dom} \; \phi$ is an interval with endpoints $\alpha<\beta$, so, $\operatorname{int} \left( \mbox{dom} \; \phi \right) =(\alpha, \beta) $. Since $\phi$ is closed, we have $\lim _{t \rightarrow \alpha_{+}} \phi(t)=\phi(\alpha)$, if $\alpha$ is finite and  $\lim _{t \rightarrow \beta_{-}} \phi(t)=\phi(\beta)$, if $\beta$ is finite.
	\smallskip
	
	We assume that  $\phi$ is nonnegative and attains its minimum at the point  $1 \in \operatorname{int}  \left(\mbox{dom} \; \phi \right)$, i.e., $\phi(1) = 0$.  The class of such functions is denoted by $\Phi$.
	\begin{definition}\label{Phi_Definition}
		Given $\phi \in \Phi$, the $\phi$-divergence of the probability measure $G$ with respect to $F$ is
		\begin{equation} \label{phi_divergence_definition}
			D_{\phi}(G\| F)=\left\{\begin{array}{cl}
				\int_{\mathcal{X}} \phi\left(\frac{g(x)}{f(x)}\right) f(x)dx & \text { if } \mathrm{G}\ll \mathrm{F} \\
				+\infty & \text { otherwise }
			\end{array}\right.
		\end{equation}
		where $f$ and $g$ are the densities of $F$ and $G$ (the distributions of $X$ with common support $\mathcal{X}$) with respect to a common dominating measure.
	\end{definition}
	To avoid pathological cases, throughout the paper, we assume the following:
	\begin{equation*}\label{Assumption_phi}
		\phi(0)<\infty, 0 \cdot \phi\left(\frac{0}{0}\right) \equiv 0,0 \cdot \phi\left(\frac{s}{0}\right)=\lim _{\varepsilon \rightarrow 0} \varepsilon  \cdot \phi\left(\frac{s}{\varepsilon}\right)=s \lim _{t \rightarrow \infty} \frac{\phi(t)}{t}, \quad s>0 .
	\end{equation*}

	It is worth remarking that the $\phi$-divergence $D_\phi(G\| F)$ is convex in both of its arguments (see Proposition \ref{phi_properties} in Appendix \ref{App_phi-div}). 
	
	A key element in our analysis is the convex conjugate of $\phi$. For $\phi \in \Phi$ its conjugate denoted by  $\phi^*$ is:
	\begin{equation}\label{Phi_convex_conjugate_new}
		\phi^*(s)=\sup _{t \in \mathbb{R}}\{s t-\phi(t)\}
	\end{equation}
	
	Two commonly used $\phi$-divergences are the Kullback--Leibler and the $\chi^2$ divergences. The Kullback--Leibler divergence corresponds to $\phi(t)=t\log t$ for $t\geq 0$, while the $\chi^2$-divergence is generated by $\phi(t)=\frac{1}{2}(t-1)^2$. Additional examples of $\phi$-divergences and their corresponding conjugate functions are provided in Table \ref{tb:div} in Appendix \ref{App_phi-div}.
	
	Endowed with the notion of $\phi$-divergence, we can reformulate the worst-case superquantile program (\ref{Robust_CVAR}) as 
	\begin{eqnarray}
		\mathbb{S}^R_\beta(Z)&=&\inf_{G\in\mathcal{U}_\phi(F,\rho)}\mathbb{S}^G_\beta(Z)\label{Robust_CVAR_phi}\\
		&=&\inf_{G\in\mathcal{U}_\phi(F,\rho)}\sup_{\lambda\in \RR}\left\{ \lambda+{1\over \beta}\EE_G[(Z-\lambda)_{-}]\right \}\nonumber
	\end{eqnarray}
	
	where $\mathcal{U}_\phi(F,\rho)$ is defined as:
	\begin{equation}\label{Uncertainty_Set_phi}
		\mathcal{U}_{\phi}(F,\rho)=\{G\ll F: D_\phi(G\| F)\leq \rho\},
	\end{equation}
	and, as above, $F$ and $G$ denote distributions for $X$ only.
	\smallskip
	
	The following lemma shows that, for a \emph{fixed} $\lambda$, the inner worst-case expectation over $\mathcal{U}_\phi(F,\rho)$ admits a tractable dual representation. This is the standard $\phi$-divergence robust-expectation duality (\cite{BentalTeboulle1987}; see also \cite{Kuhn_Shafiee_Wiesemann_2025}), and requires no exchange of $\inf$ and $\sup$.
	
	\begin{lemma}\label{phi_dual_lemma} Let $\phi\in \Phi$, $\rho>0$, and let $h:\mathcal{X}\to\RR$ be measurable and bounded below. Then
		\begin{equation}\label{Inner_Dual}
			\inf_{G\in\mathcal{U}_\phi(F,\rho)}\EE_G[h(X)]=\sup_{\eta\in \RR,\,\kappa\geq 0}\left\{\eta-\kappa\rho-\kappa\,\EE_F\left[\phi^*\left({\eta-h(X)\over \kappa}\right)\right]\right\},
		\end{equation}
		with the convention that the $\kappa=0$ term is evaluated via the recession function of $\phi^*$.
	\end{lemma}
	\proof All proofs are collected in Appendix \ref{Sec:Proofs_WDZ} \eproof
	
	Applying Lemma \ref{phi_dual_lemma} with $h(X)=(Z-\lambda)_-$ and substituting into \eqref{Robust_CVAR_phi}, we obtain the following representation of the robust superquantile.
	
	\begin{proposition}\label{Robust_CVAR_Dual_Prop} Let $\phi\in \Phi$ and $\beta\in (0,1)$. Then 
		\begin{equation}\label{Robust_CVAR_Reformulation}
			\mathbb{S}^R_\beta(Z)=\sup_{\lambda,\eta \in \RR,\ \kappa\geq 0}\left\{ \lambda+{1\over \beta}\left[\eta-\kappa\rho-\kappa\,\EE_F\left[\phi^{*}\left({\eta-(Z-\lambda)_-\over \kappa}\right) \right]\right]\right\}.
		\end{equation}
		The objective in (\ref{Robust_CVAR_Reformulation}) is jointly concave in $(\lambda,\eta,\kappa)$, so $\mathbb{S}^R_\beta(Z)$ is the value of a finite-dimensional concave program.
	\end{proposition}

	Equation (\ref{Robust_CVAR_Reformulation}) is the object we use to compute $\mathbb{S}^R_\beta(Z)$ in Section \ref{Sec:Computing_Robust} below. Note that $\rho=0$ forces $\eta=h(X)$ a.e.\ $F$ for the inner program to be finite, which collapses (\ref{Robust_CVAR_Reformulation}) back to the original (non-robust) variational definition (\ref{Lower_CVAR_Equation_Def}); we record this as part of Remark \ref{Remark_Kallus_Nesting} below.
	\smallskip
	
	A tractable parametric family within $\Phi$ is the Cressie--Read family,
	\begin{equation}\label{Cressie_Read}
		\phi_{cr_k}(t):=\frac{t^k-k t+k-1}{k(k-1)}, \qquad \phi_{cr_k}^*(s):=\frac{1}{k}\left[((k-1) s+1)_{+}^{k_*}-1\right], \qquad k_*=\frac{k}{k-1},
	\end{equation}
	which nests the $\chi^2$-divergence at $k=2$ and (in the limit $k\to1$) the Kullback--Leibler divergence. Substituting $\phi^*_{cr_k}$ into (\ref{Robust_CVAR_Reformulation}) yields a concave program in $(\lambda,\eta,\kappa)$ with a closed-form integrand, which we use for estimation in Section \ref{Sec:Computing_Robust}.

	\section{Bounds}\label{Sec:Bounds_Ind_Welfare}
	In this section, we study the distribution of individual welfare changes without imposing strong structural assumptions or requiring detailed micro-level data. We focus on the empirically familiar setting in which only aggregate market data is available and demonstrate that the same information used to identify average welfare changes can, through superquantiles, yield informative insights into the distributional consequences of policy interventions.

	\subsection{Bounding individual welfare changes}
	We begin by defining the individual welfare change as the random variable 
	\begin{equation}\label{delta_u_def}
		\tau(X,t,\varepsilon)\triangleq \max_{y\in B}{U(y;X,t_1,\varepsilon)}-\max_{y\in B}{U(y;X,t_0,\varepsilon)},
	\end{equation}
	where $X \in \mathcal{X}$ denotes observable characteristics, and $t = (t_0, t_1)$ represents the values of a predetermined variable before and after an intervention. For example, $t_0$ and $t_1$ may correspond to prices observed before and after a policy intervention, such as the introduction of an ad valorem tax. Similarly, $t$ indicates whether an individual has been assigned to a social program: $t_1=1$ and $t_0=0$, respectively.
	
	\smallskip
	
	For a type $(X,\varepsilon)$, the realization of the random variable $\tau(X,t,\varepsilon)$ is generally unobservable to the analyst. More importantly, in many relevant economic settings, the distribution of $\tau(X,t,\varepsilon)$ is unknown, posing two key identification challenges when analyzing the distribution of individual welfare. The first arises in the typical situation in which the analyst observes only aggregate market data. Because individual decision-making is not available, the conventional approach in such settings is to work with the conditional average welfare change, defined as
	\begin{eqnarray}\label{CATE_delta_u}
		\tau(X,t)&\triangleq& \EE\left(\tau(X,t,\varepsilon)\mid X,t\right),\\
		&=&\EE\left(\max_{y\in B}U(y;X,t_1,\varepsilon)\mid X, t_1\right)-\EE\left(\max_{y\in B}U(y;X,t_0,\varepsilon)\mid X, t_0\right)\nonumber
	\end{eqnarray}
	where the expectations are taken with respect to $\varepsilon$.\footnote{We note that the second equality uses the assumption that $F_{\varepsilon\mid X,t_1}=F_{\varepsilon\mid X,t_0}=F_{\varepsilon\mid X}$ for all $X\in \mathcal{X}$.}
	\smallskip
	
	As shown in \citet{Allen_Rehbeck}, $\tau(X,t)$ is non-parametrically identified for general distributions of $\varepsilon$ and flexible specifications of the deterministic utility component $u(X)$. However, such identification results are typically uninformative about the distributional implications of policy interventions.\footnote{See \cite{bhattacharya2024nonparametric} for an excellent survey of approaches to achieve identification of welfare changes in demand models with aggregate and micro-data.}
	\smallskip

	A second issue may arise even when the analyst has access to rich micro-level data: identifying $\tau(X,t,\varepsilon)$ remains impossible due to the evaluation problem. To illustrate, suppose $t_1$ indicates that a worker participates in a training program aimed at increasing productivity, while $t_0$ represents the status quo of non-participation. In this environment, each worker is in only one of these states at a time, but never both. The treatment effect literature has traditionally addressed this challenge by also focusing on \emph{average} treatment effects, which again does not address the distributional implications of interventions \citep{HECKMAN2007_Handbook_partI}.
	
	\smallskip

	Therefore, while directly identifying $\tau(X,t,\varepsilon)$ remains infeasible in the two scenarios above, our goal is to gain insight into the distributional behavior of this random variable by exploiting the information in $\tau(X,t)$. For this purpose, we define the $\beta$-superquantile  associated with  $\tau(X,t,\varepsilon)$ as follows:
	\begin{equation}\label{delta_w_pop}
		\mathbb{S}_\beta(\tau(X,t,\varepsilon))\triangleq \mathbb{S}_\beta\Big(\max_{y\in B}U(y;X,t_1,\varepsilon)-\max_{y\in B}U(y;X,t_0,\varepsilon)\Big)
	\end{equation}
	
	Expression (\ref{delta_w_pop}) is concerned with the left tail of  $\tau(X,t,\varepsilon)$, which depends 
	on the joint distribution of $X$ and $\varepsilon$. In words, $\mathbb{S}_\beta(\tau(X,t,\varepsilon))$ quantifies the average change in utility for the bottom ($100\times\beta)\%$ of the population affected by an intervention that changes $t_0$ to $t_1$. From an economic standpoint, expression (\ref{delta_w_pop}) can be interpreted as a welfare measure of how the bottom ($100\times\beta)\%$ of all individuals fare from this intervention.\footnote{It is worth noting that the superquantile functional $\mathbb{S}_\beta(\tau(X,t,\varepsilon))$ is not additive. As a result $\mathbb{S}_\beta(\tau(X,t,\varepsilon))\neq \mathbb{S}_\beta\Big(\max_{y\in B}U(y;X,t_1,\varepsilon)\Big)-\mathbb{S}_\beta\Big(\max_{y\in B}U(y;X,t_0,\varepsilon)\Big)$. Consequently, differences in superquantiles of indirect utility cannot be used to infer the distribution of individual welfare changes.}
	\smallskip

	The following theorem provides an upper bound for Eq.~(\ref{delta_w_pop}) that does not rely on parametric assumptions nor detailed individual-level data.

	\begin{theorem}\label{Bound_Change_Utility} Let Assumption \ref{Assumption_PUM1} hold. 
		Then for all $\beta\in(0,1]$:
		\begin{equation}\label{Aggregate_Bound_Change_W}
			\mathbb{S}_\beta(\tau(X,t,\varepsilon)) \leqslant \mathbb{S}_\beta(\tau(X,t)).
		\end{equation}
	\end{theorem}

	Theorem \ref{Bound_Change_Utility} establishes that the superquantile of individual welfare changes is bounded above by the superquantile of conditional average welfare changes.\footnote{Theorem \ref{Bound_Change_Utility} uses only conditional means, and in general it is not sharp. When the conditional distributions given $X$ of the welfare levels before and after the intervention are separately identified---as for the potential outcomes $(V_0,V_1)$ in the generalized Roy model of Section \ref{Sec:Gen_Roy_Cost}, where the conditioning may also include $U_{\mathcal{W}}$---the bound can be sharpened. By \citet[Thm.~3.2]{Fan_Guerre_Zhu_2017}, combined with the variational formula (\ref{Lower_CVAR_Equation_Def}), the sharp upper and lower bounds on the superquantile of individual changes are attained when the two levels are conditionally comonotone (rank invariance) and conditionally countermonotone (rank reversal), respectively. We work with the conditional-mean bound because it requires only aggregate data.} The significance of bound (\ref{Aggregate_Bound_Change_W}) lies in its empirical feasibility. In particular, the left-hand side cannot be estimated without detailed individual-level data and/or strong parametric assumptions. In contrast, the right-hand side can be identified non-parametrically from aggregate data alone. For example, identification can be achieved within the PUM framework using the results of \citet{Allen_Rehbeck}, or under the class of RUMs following the nonparametric approach of \citet{Berry_Haile_ECMA2014}.  Similarly, bound \eqref{Aggregate_Bound_Change_W} is also relevant when the analyst faces missing outcomes, as discussed above, which persists even with rich micro-level data.\footnote{In Section \ref{Sec:Gen_Roy_Cost} we discuss this problem in further detail.}
	\smallskip
	
	From an economic perspective, Theorem \ref{Bound_Change_Utility} is valuable because it offers a framework for evaluating the distributional welfare effects of policy interventions. To illustrate, define the total average welfare change as $\tau(t) = \EE_{F_{X}}(\tau(X,t))$. Suppose a policy yields $\tau(t) > 0$, but $\mathbb{S}_\beta(\tau(X,t)) < 0$ for a substantial fraction of the population (e.g., $\beta = 0.2$). Comparing these two quantities provides a first-order assessment of whether a policy generates overall welfare gains while imposing losses on specific subgroups. This information might indicate that the policy change yields some large ``winners" from the policy, masking harm that other subpopulations may be experiencing. Furthermore, $\mathbb{S}_\beta(\tau(X,t))$ enables us to identify which subpopulations are adversely affected. Assuming continuity, $\mathbb{S}_\beta(\tau(X,t))$ corresponds to the average welfare change among agents with $\tau(X,t) \leq F^{-1}_{\tau(X,t)}(\beta)$. Since $\tau(X,t)$ is identifiable, this approach allows us to identify which observable subgroups experience welfare losses under the policy. It also provides insights into how to improve a policy and minimize these harms.  
	\smallskip
	
	The following result is a direct corollary of  Theorem \ref{Bound_Change_Utility}.
	\begin{corollary}\label{Quant_Aggregate_Bound_Change_W}Let Assumption \ref{Assumption_PUM1} hold. In addition, assume that $\tau(X,t,\varepsilon)$ and $\tau(X,t)$ have continuous distributions. Then the bound (\ref{Aggregate_Bound_Change_W}) can be equivalently written as:
		\begin{equation}\label{Quan_bound}
			\int_0^\beta F^{-1}_{\tau(X,t,\varepsilon)}(\theta)d\theta\leq  \int_0^\beta F^{-1}_{\tau(X,t)}(\theta)d\theta 
		\end{equation}
		
	\end{corollary}
	
	It is worth remarking that $F^{-1}_{\tau(X,t,\varepsilon)}(\theta)$ corresponds to the $\theta$-quantile of the individual utility changes. In general, as we said earlier, this quantity is not identified. On the other hand,  $F^{-1}_{\tau(X,t)}(\theta)$ gives us the $\theta$-quantile of the conditional average welfare changes, as measured by $\tau(X,t)$.  
	\smallskip

	One final point worth mentioning regarding Theorem \ref{Bound_Change_Utility}: We did not exploit any structure of the PUM framework in its derivation, so the bound is general and applies beyond the additive structure of the PUM class; indeed, as discussed in Section \ref{sec:lit} and Remark \ref{Remark_Kallus_Nesting}, the inequality (\ref{Aggregate_Bound_Change_W}) itself is mathematically equivalent to Theorem 3.1 in \cite{Kallus2023}. Still, PUM is the most general class we know of for which the bound can be non-parametrically estimated using only aggregate data; with rich micro-level data, identification of $\tau(X,t)$ is also possible if utilities are additive (see, e.g., \cite{Bhattacharya_2015,bhattacharya2024nonparametric}) or under certain conditions if utilities are non-additive (see e.g., \cite{imbens2009identification}). Section \ref{Sec:Robust_Bounds} below extends Theorem \ref{Bound_Change_Utility} to allow for ambiguity in the distribution of $X$, which is where the genuinely new theoretical content of this part of the paper lies.
	\smallskip
	
	Theorem \ref{Bound_Change_Utility} holds for every $\beta\in(0,1]$ simultaneously, and this uniformity has a sharper implication than any single-$\beta$ statement on its own.
	
	\begin{corollary}[Superquantile dominance implies second-order stochastic dominance]\label{Cor:SSD}
		Let Assumption \ref{Assumption_PUM1} hold. Then $\tau(X,t)$ second-order stochastically dominates $\tau(X,t,\varepsilon)$: for every nondecreasing, concave function $u:\RR\to\RR$,
		\begin{equation}\label{SSD_eq}
			\EE\big[u(\tau(X,t,\varepsilon))\big] \;\leq\; \EE\big[u(\tau(X,t))\big].
		\end{equation}
	\end{corollary}
	
	\begin{remark}
		Corollary \ref{Cor:SSD} strengthens Theorem \ref{Bound_Change_Utility} from a single-$\beta$ statement into a distribution-wide ranking: because $\mathbb{S}_\beta(\tau(X,t,\varepsilon))\leq\mathbb{S}_\beta(\tau(X,t))$ holds for \emph{every} $\beta\in(0,1]$ at once, it implies that every risk-averse expected-utility-maximizing planner---not only one evaluating a specific tail fraction $\beta$---weakly prefers the conditional-mean welfare change $\tau(X,t)$ to the individual welfare change $\tau(X,t,\varepsilon)$. Formally, uniform-in-$\beta$ superquantile dominance is equivalent to second-order stochastic dominance \citep[Thm.~3.1]{OgryczakRuszczynski2002}, via the identity $\beta\,\mathbb{S}_\beta(Z)=\int_0^\beta F_Z^{-1}(\theta)\,d\theta$ relating the superquantile to the absolute Lorenz function of $Z$.
	\end{remark}
	\smallskip

	The following proposition provides a lower bound on the buffered probability of individual welfare loss.
	\begin{proposition}\label{Prop:bPOE}
		Let Assumption \ref{Assumption_PUM1} hold, and fix a threshold $\theta$ in the common domain of $p_\theta(\tau(X,t,\varepsilon))$ and $p_\theta(\tau(X,t))$. Then
		\begin{equation}\label{bPOE_bound}
			p_\theta(\tau(X,t))\;\leq\;p_\theta(\tau(X,t,\varepsilon)).
		\end{equation}
	\end{proposition}
	\smallskip

	From a policy standpoint, Proposition~\ref{Prop:bPOE} addresses a different question from Theorem~\ref{Bound_Change_Utility}. Theorem~\ref{Bound_Change_Utility} fixes a tail fraction $\beta$ in advance---for example, the worst $20\%$ of the population---and bounds the average welfare change within that group. In many applications, however, the relevant object is instead an exogenously specified threshold, such as a benefit cliff, a poverty line, or a dollar loss that triggers eligibility for compensation, regulatory review, or legal action. Proposition~\ref{Prop:bPOE} addresses precisely this setting. For a given threshold $\theta$, it guarantees that at least $p_\theta(\tau(X,t))\times 100\%$ of the population experiences a welfare change no greater than $\theta$, using only the identified conditional average welfare change. Thus, the analyst need not specify a tail fraction $\beta$ in advance or observe individual-level welfare changes. The proposition therefore provides a natural statistic for regulatory or policy-making tests, where the relevant question is how many individuals experience a loss of at least a specified magnitude, rather than the average loss among a prespecified fraction of the population.	\smallskip
	\smallskip

	This inversion also distinguishes Proposition~\ref{Prop:bPOE} from the treatment-effect-risk bounds in \citet{Kallus2023}. At the level of the basic, non-robust bound, Theorem~\ref{Bound_Change_Utility} is mathematically equivalent to his Theorem 3.1. However, \citet{Kallus2023} states the resulting bounds in superquantile-magnitude form and does not provide an analog of Proposition \ref{Prop:bPOE}, which inverts the bound to obtain a lower bound on the fraction of the population whose realized welfare change falls below a given threshold. More broadly, buffered probabilities of exceedance have been used in portfolio optimization, structural reliability, and natural-hazard risk \citep{Rockafellar_Royset_2010,Mafusalov_Uryasev_2018}. To our knowledge, Proposition~\ref{Prop:bPOE} provides the first application of this approach to treatment-effect or welfare analysis.
	\smallskip

	We next derive a lower bound for $\mathbb{S}_\beta(\tau(X,t,\varepsilon))$. The following theorem establishes such a bound under an additional condition.	
	\begin{theorem}\label{Bound_Change_Utility_Lower} Let Assumption \ref{Assumption_PUM1} hold. In addition, suppose that $\tau(X,t)-\tau(X,t,\varepsilon)\leq \gamma$  for $X\in\mathcal{X}$ and $\gamma\in \RR$. Then for all $\beta\in (0,1]$: 
		\begin{equation}\label{Aggregate_Bound_Change_W_Lower}
			\mathbb{S}_\beta(\tau(X,t,\varepsilon))\geq \mathbb{S}_\beta(\tau(X,t))-\gamma.
		\end{equation}
	\end{theorem}
	
	The preceding result shows that if the difference between $\tau(X,t,\varepsilon)$ and $\tau(X,t)$ is uniformly bounded by a known constant $\gamma$, then
	$$	\mathbb{S}\beta(\tau(X,t,\varepsilon))
	\geq
	\mathbb{S}\beta(\tau(X,t))-\gamma.$$
	The right-hand side is identified given $\gamma$; see, for example, \citet{Manski1990} on identification under bounded-support restrictions.
	
	\smallskip
	
	Theorems~\ref{Bound_Change_Utility} and~\ref{Bound_Change_Utility_Lower} differ in the information they require. Theorem~\ref{Bound_Change_Utility} requires neither knowledge of the distribution of unobservables nor parametric restrictions on that distribution. By contrast, Theorem~\ref{Bound_Change_Utility_Lower} imposes a restriction on the range of individual welfare changes conditional on observables. To illustrate the distinction, consider the special case in which the PUM reduces to a standard ARUM. In this setting, \citet{bhattacharya2024nonparametric} shows that average welfare and observed choice probabilities do not identify the distribution of individual welfare changes. Our results establish a complementary point: identified average welfare changes can nevertheless provide informative bounds on features of that distribution. The distinction arises from the information used. Our bounds exploit the identified conditional average welfare change directly, whereas the bounds based solely on observed choice probabilities do not incorporate this additional welfare information.
	
	\subsection{Computing $\mathbb{S}_\beta(\tau(X,t))$}\label{sec{computin_LP}} We now discuss a simple approach to computing $\mathbb{S}_\beta(\tau(X,t))$. Suppose the analyst has access to a collection of $K$ estimates $\hat{\tau}(X_1), \ldots, \hat{\tau}(X_K)$ of the conditional means at $K$ covariate cells with known weights $w_k>0$, $\sum_{k=1}^Kw_k=1$ (equal weights $w_k=1/K$ correspond to the empirical distribution of $K$ draws of $X$; see Remark~\ref{Rem:Weights}), where, for ease of exposition, we omit the dependence on $t$. Using these estimates, a direct approach to computing $\mathbb{S}_\beta(\tau(X,t))$ is to employ the variational representation \eqref{Optimal_value_S} together with a plug-in method, yielding the following program:
	\begin{equation}\label{S_estimates}
		\hat{\mathbb{S}}_\beta(\hat{\tau}(X,t))
		= \max\limits_{\lambda \in \mathbb{R}}
		\left\{
		\lambda + \frac{1}{\beta} \sum_{k=1}^K w_k
		\min\{\hat{\tau}(X_k) - \lambda, 0\}
		\right\}
	\end{equation}

	Problem \eqref{S_estimates} is concave but, unfortunately, non-smooth in $\lambda$. However, and following \cite{Rockafellar2000OptimizationOC}, we can avoid the analytical complications introduced by non-smoothness by observing that program \eqref{S_estimates} can be equivalently expressed as a linear programming problem:
	\begin{eqnarray}
		V(\hat{\lambda})=\max_{\lambda,z}&&\left\{\lambda+{1\over\beta}\sum_{k=1}^K w_kz_k\right\},\label{LP_Superquantile}\\
		\text{subject to:}&&\hat{\tau}(X_k)-\lambda\geq z_k,\text{for \quad$k=1,\ldots,K$}\nonumber\\
		&&z_k\leq 0.\nonumber
	\end{eqnarray}
	
	Program \eqref{LP_Superquantile} offers a straightforward way to compute $\hat{\lambda}$ and $\hat{\mathbb{S}}_\beta(\hat{\tau}(X,t))$. The problem involves a linear objective function with only $2K$ linear constraints, making it computationally simple and easy to implement with standard optimization software.
	
	The value of $\hat{\lambda}$ can be interpreted as an estimator of a $\beta$-quantile of the random variable $\tau(X,t)$ (the set of optimal $\lambda$ is an interval when the cumulative weight of the lowest cells equals $\beta$ exactly, for instance when $\beta K$ is an integer under equal weights), while the value of the objective function at the optimum, $V(\hat{\lambda})$, provides an estimate of $\mathbb{S}_\beta(\tau(X,t))$.

	Because this approach takes the pre-estimated values $\hat{\tau}(X_1),\ldots,\hat{\tau}(X_K)$ as inputs, it is a linear program with estimated inputs, and sampling variability must be accounted for in inference. The dual of \eqref{LP_Superquantile} is $\min_{\mu\in M_\beta}\sum_k\mu_k\hat\tau(X_k)$ with $M_\beta=\{\mu:\,0\le\mu_k\le w_k/\beta,\ \sum_k\mu_k=1\}$, the sample counterpart of the risk-envelope representation underlying Remark~\ref{Remark_Kallus_Nesting}, where $\mu_k$ is the weight that cell $k$ receives in the worst-$\beta$ tail. Two features matter for inference. When the cumulative weight of the lowest cells equals $\beta$ exactly, the optimal $\lambda$ is a whole interval rather than a point, and the value $\hat{\mathbb{S}}_\beta$, although always unique, is not differentiable in $\hat\tau$ whenever cells with different tail status (fully in the tail, on its boundary, or outside it) tie. Pointwise delta-method arguments are therefore not uniformly valid. Appendix~\ref{App:LP_Inference} gives a confidence interval that remains valid at such ties, obtained by projecting a confidence set for the parameters governing $\tau(X,t)$, treating the cells $X_1,\ldots,X_K$ as fixed; when the $\hat\tau(X_k)$ have a nonsingular joint covariance, the primal--dual confidence sets of \cite{Shum_2022} are an alternative.

	\subsection{Robust bounds on individual welfare changes}\label{Sec:Robust_Bounds} We now show that the bounds established above remain valid under ambiguity about the distribution of $X$, using the distributionally robust framework introduced in Section~\ref{Sec:DRO_Framework}. Throughout this section, $\tau(X,t,\varepsilon)$ and
	$$
	\tau(X,t)=\EE_F[\tau(X,t,\varepsilon)\mid X]
	$$
	are defined as in (\ref{delta_u_def})--(\ref{CATE_delta_u}). The uncertainty set $\mathcal{U}_\phi(F,\rho)$ in (\ref{Uncertainty_Set_phi}) perturbs only the marginal distribution of $X$, while holding the conditional distribution $F_{\varepsilon\mid X}$ fixed at its nominal value. Consequently, the function $\tau(\cdot,t)$ is invariant across distributions in $\mathcal{U}_\phi(F,\rho)$. For each $G\in\mathcal{U}_\phi(F,\rho)$, let $\mathbb{S}^G_\beta(\tau(X,t,\varepsilon))$ denote the superquantile of $\tau(X,t,\varepsilon)$ under the joint distribution $G\otimes F_{\varepsilon\mid X}$.

	\begin{theorem}\label{Robust_Bound_Change_Utility} Let Assumption \ref{Assumption_PUM1} hold. Then for all $\beta\in (0,1]$ and  any $\phi\in\Phi$  and $\rho>0$,
		\begin{equation}\label{Robust_Aggregate_Bound_Change_W}
			\mathbb{S}_\beta^R(\tau(X,t,\varepsilon))\leqslant \mathbb{S}_\beta^R(\tau(X,t)).
		\end{equation}
	\end{theorem}

	\begin{remark}\label{Remark_Kallus_Nesting}
		Theorem~\ref{Robust_Bound_Change_Utility} nests Theorem~\ref{Bound_Change_Utility} when $\rho=0$. In this case, $\mathcal{U}_\phi(F,0)=\{F\}$, and (\ref{Robust_Aggregate_Bound_Change_W}) reduces to (\ref{Aggregate_Bound_Change_W}). As discussed in Section~\ref{sec:lit}, this result is mathematically equivalent to Theorem 3.1 of \citet{Kallus2023}, formulated there in terms of the conditional average treatment effect rather than $\tau(X,t)$; we therefore do not claim novelty for the case $\rho=0$. The contribution of Theorem \ref{Robust_Bound_Change_Utility} is to extend this result to $\rho>0$. To the best of our knowledge, this extension is new: it establishes that the upper bound on the superquantile of individual welfare changes continues to hold when the marginal distribution of $X$ is itself ambiguous, with the extent of this ambiguity controlled by $\rho$.
	\end{remark}
	
	Proposition \ref{Prop:bPOE}'s probabilistic reading extends to the robust bound in the same way.
	
	\begin{corollary}\label{Cor:bPOE_Robust}
		Let Assumption \ref{Assumption_PUM1} hold. Then for any $\phi\in\Phi$, $\rho>0$, and threshold $\theta$ in the common domain,
		\begin{equation}\label{bPOE_bound_robust}
			p_\theta^R(\tau(X,t);\rho)\;\leq\;p_\theta^R(\tau(X,t,\varepsilon);\rho),
		\end{equation}
		where $p_\theta^R(Z;\rho)\triangleq\inf\{\beta\in(0,1]:\mathbb{S}_\beta^R(Z)\geq \theta\}$ inverts the robust superquantile of Theorem \ref{Robust_Bound_Change_Utility}.
	\end{corollary}
	
	Corollary~\ref{Cor:bPOE_Robust} extends the probabilistic interpretation of Proposition~\ref{Prop:bPOE} to settings in which the distribution of $X$ is ambiguous. In particular, $p_\theta^R(\tau(X,t);\rho)$, computable from the identified conditional average welfare change alone, provides a valid lower bound on the buffered probability of individual welfare loss uniformly over all distributions of $X$ in the $\phi$-divergence neighborhood $\mathcal{U}_\phi(F,\rho)$ of the nominal distribution $F$.
	\smallskip
	
	The same logic extends the lower bound of Theorem \ref{Bound_Change_Utility_Lower}.
	
	\begin{theorem}\label{Robust_Bound_Change_Utility_Lower} Let Assumption \ref{Assumption_PUM1} hold. In addition, suppose that $\tau(X,t)-\tau(X,t,\varepsilon)\leq\gamma$ for $X\in\mathcal{X}$ and $\gamma\in\RR$. Then for all $\beta\in(0,1]$ and any $\phi\in\Phi$ and $\rho>0$,
		\begin{equation}\label{Robust_Aggregate_Bound_Change_W_Lower}
			\mathbb{S}_\beta^R(\tau(X,t,\varepsilon))\geq \mathbb{S}_\beta^R(\tau(X,t))-\gamma.
		\end{equation}
	\end{theorem}

	\begin{remark}
		As in Remark~\ref{Remark_Kallus_Nesting}, Theorem~\ref{Robust_Bound_Change_Utility_Lower} reduces to Theorem~\ref{Bound_Change_Utility_Lower} when $\rho=0$. As discussed in Section~\ref{sec:lit}, the latter is mathematically equivalent to Theorem~3.3 of \citet{Kallus2023}, formulated here in terms of individual welfare changes. Importantly, extending the result to $\rho>0$ does not require strengthening the range restriction involving $\gamma$: the same constant $\gamma$ that bounds $\tau(X,t)-\tau(X,t,\varepsilon)$ in the non-robust case continues to yield a valid lower bound uniformly over all distributions of $X$ in $\mathcal{U}_\phi(F,\rho)$.
	\end{remark}
	\subsection{Computing $\mathbb{S}_\beta^R(\tau(X,t))$}\label{Sec:Computing_Robust} We end this section by turning to computing the robust upper bound $\mathbb{S}_\beta^R(\tau(X,t))$ in (\ref{Robust_Aggregate_Bound_Change_W}), paralleling the plug-in/linear-programming approach above for the non-robust bound. As before, suppose the analyst has access to a collection of $K$ estimates $\hat\tau(X_1),\ldots,\hat\tau(X_K)$. Replacing $F$ in (\ref{Robust_CVAR_Reformulation}) with the $K$-point distribution $F_K$ that places weight $w_k$ on each $\hat\tau(X_k)$ (weight $1/K$ for the empirical distribution) yields the sample analogue
	\begin{equation}\label{Robust_S_estimates}
		\hat{\mathbb{S}}^R_\beta(\hat\tau(X,t))=\max_{\lambda,\eta\in\RR,\ \kappa\geq0}\left\{\lambda+\frac1\beta\left[\eta-\kappa\rho-\kappa\sum_{k=1}^K w_k\,\phi^*\left(\frac{\eta-(\hat\tau(X_k)-\lambda)_-}{\kappa}\right)\right]\right\}.
	\end{equation}
	Program (\ref{Robust_S_estimates}) is a finite-dimensional concave maximization in $(\lambda,\eta,\kappa)$ and can be solved with standard convex optimization software; under the Cressie--Read specialization (\ref{Cressie_Read}) the integrand $\phi^*_{cr_k}$ is available in closed form, which we exploit in the simulations of Section \ref{Sec:Simulations} below. As $\rho\to0$, (\ref{Robust_S_estimates}) recovers the linear program (\ref{LP_Superquantile}) for the non-robust bound, consistent with Remark \ref{Remark_Kallus_Nesting}.
	
	\begin{remark}[Weights]\label{Rem:Weights}
		The weights $w_k$ are the masses that the nominal distribution $F$ of $X$ places on the $K$ cells, $w_k=\Pr_F(X\in\text{cell }k)$; they enter because $\mathbb{S}_\beta(\tau(X,t))$ is a functional of the law of $X$ and describes the worst-off $\beta$-share of the \emph{population}, not of the cells. Equal weights $w_k=1/K$ are appropriate when each cell is a single observation (so that $F_K$ is the empirical distribution) or when cells are equally populated, for instance, quantile bins. Otherwise, the $w_k$ may be sample shares $n_k/n$, shares of a target population to which the policy applies, or survey weights. In the last two cases, and whenever the weights are estimated, the nominal $w$ is only a best guess of the true distribution of $X$, and this is precisely the ambiguity that the uncertainty set $\mathcal{U}_\phi(F,\rho)$ of Section~\ref{Sec:DRO_Framework} is designed to absorb. For sample shares and the $\chi^2$ member of \eqref{Cressie_Read} ($k=2$), $n\,D_\phi(G\|F)$ is asymptotically $\tfrac12\chi^2_{K-1}$ at the true $G$, so that $\rho=\chi^2_{K-1}(1-\alpha)/(2n)$ yields an ambiguity set that contains the true weights with asymptotic probability $1-\alpha$.
	\end{remark}
	
	\section{Simulations}\label{Sec:Simulations}
	
	This section provides a numerical illustration of Theorems~\ref{Bound_Change_Utility}--\ref{Robust_Bound_Change_Utility_Lower}. We construct a discrete-choice environment with heterogeneous observables and simulate individual-level welfare changes. This allows us to compare the latent superquantile of individual welfare changes with the bounds derived above, which are computable using only conditional average welfare changes.
	
	\subsection{Design}\label{Sec:Sim_Design} We consider $K=8$ co-variate cells (``markets''), indexed $k=1,\ldots,K$, each with its own deterministic utility vector over $J=3$ alternatives, before ($t_0$) and after ($t_1$) an intervention, and a population weight, $w_k$, summing to one across markets. For each market we simulate $S=200{,}000$ individuals with i.i.d.\ draws of the unobserved heterogeneity $\varepsilon_i$, using \emph{the same draw} $\varepsilon_i$ for both $t_0$ and $t_1$ so that the individual pairing defining $\tau(X,t,\varepsilon)$ in (\ref{delta_u_def}) is preserved.\footnote{This is a deliberate design choice: drawing $\varepsilon$ independently across $t_0$ and $t_1$ would instead simulate the difference of two superquantiles of utility \emph{levels}, $\mathbb{S}_\beta(\max_y U(y;X,t_1,\varepsilon))-\mathbb{S}_\beta(\max_y U(y;X,t_0,\varepsilon))$, which footnote 5 notes is not equal to $\mathbb{S}_\beta(\tau(X,t,\varepsilon))$ because the superquantile functional is not additive.} For each individual this yields the exact realization $\tau_i^{(k)}=\max_y U(y;X_k,t_1,\varepsilon_i)-\max_y U(y;X_k,t_0,\varepsilon_i)$, and averaging within market $k$ yields the oracle conditional mean $\tau(X_k,t)$, the quantity that plays the role of $\hat\tau(X_k)$ in (\ref{S_estimates}) and (\ref{Robust_S_estimates}), with the population weights $w_k$ entering as in those programs. We report results under both a logit specification of $\varepsilon$ (independent Type-I extreme value draws) and a probit specification (correlated Gaussian draws with a Toeplitz covariance structure), to verify the bounds are not an artifact of either distributional choice.
	
	\subsection{Verifying Theorems \ref{Bound_Change_Utility} and \ref{Bound_Change_Utility_Lower}}\label{Sec:Sim_Layer1}
	
\begin{figure}[ht]
		\centering
		\includegraphics[width=\textwidth]{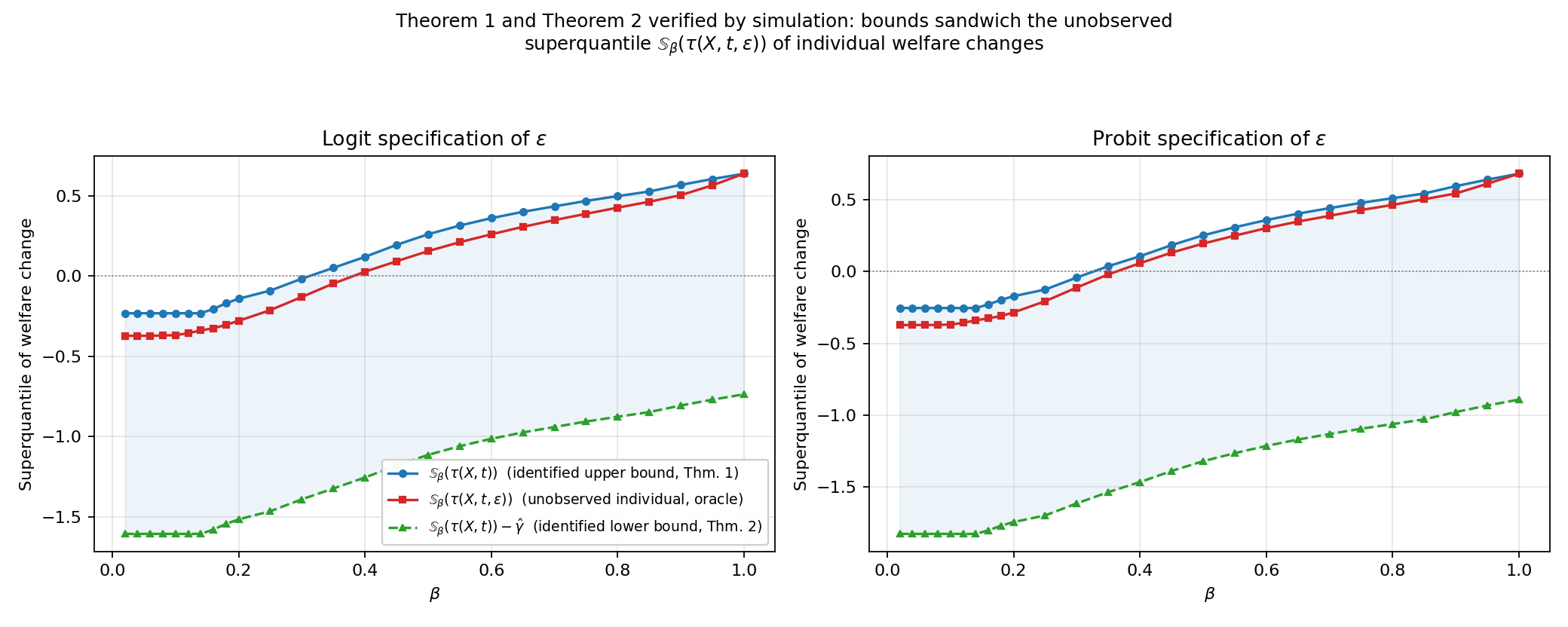}
		\caption{Theorem \ref{Bound_Change_Utility} (upper bound) and Theorem \ref{Bound_Change_Utility_Lower} (lower bound) verified by simulation. The unobserved superquantile of individual welfare changes (red) is sandwiched between the identified upper bound (blue) and lower bound (green), for both a logit and a probit specification of $\varepsilon$, across $\beta\in(0,1]$.}
		\label{fig:layer1_main}
	\end{figure}

	Figure \ref{fig:layer1_main} plots, across a grid of $\beta$, the pooled (population-weighted) superquantile of the simulated individual changes $\tau_i^{(k)}$ against the superquantile of the $K$ oracle conditional means $\tau(X_k,t)$, together with the Theorem \ref{Bound_Change_Utility_Lower} lower bound $\mathbb{S}_\beta(\tau(X,t))-\hat\gamma$, where $\hat\gamma$ is the plug-in estimate $\hat\gamma=\max_k\big[\tau(X_k,t)-\min_i \tau_i^{(k)}\big]$. Both inequalities hold with zero violations across the full grid of $\beta\in(0,1]$ and under both specifications of $\varepsilon$. More importantly, Theorem \ref{Bound_Change_Utility}'s upper bound is informative: the gap between $\mathbb{S}_\beta(\tau(X,t,\varepsilon))$ and $\mathbb{S}_\beta(\tau(X,t))$ is on the order of $0.05$ to $0.14$ across the values of $\beta$ we consider, not merely valid but practically useful for learning about the worst-affected subpopulations from conditional-average data alone.
	
\begin{figure}[ht]
		\centering
		\includegraphics[width=\textwidth]{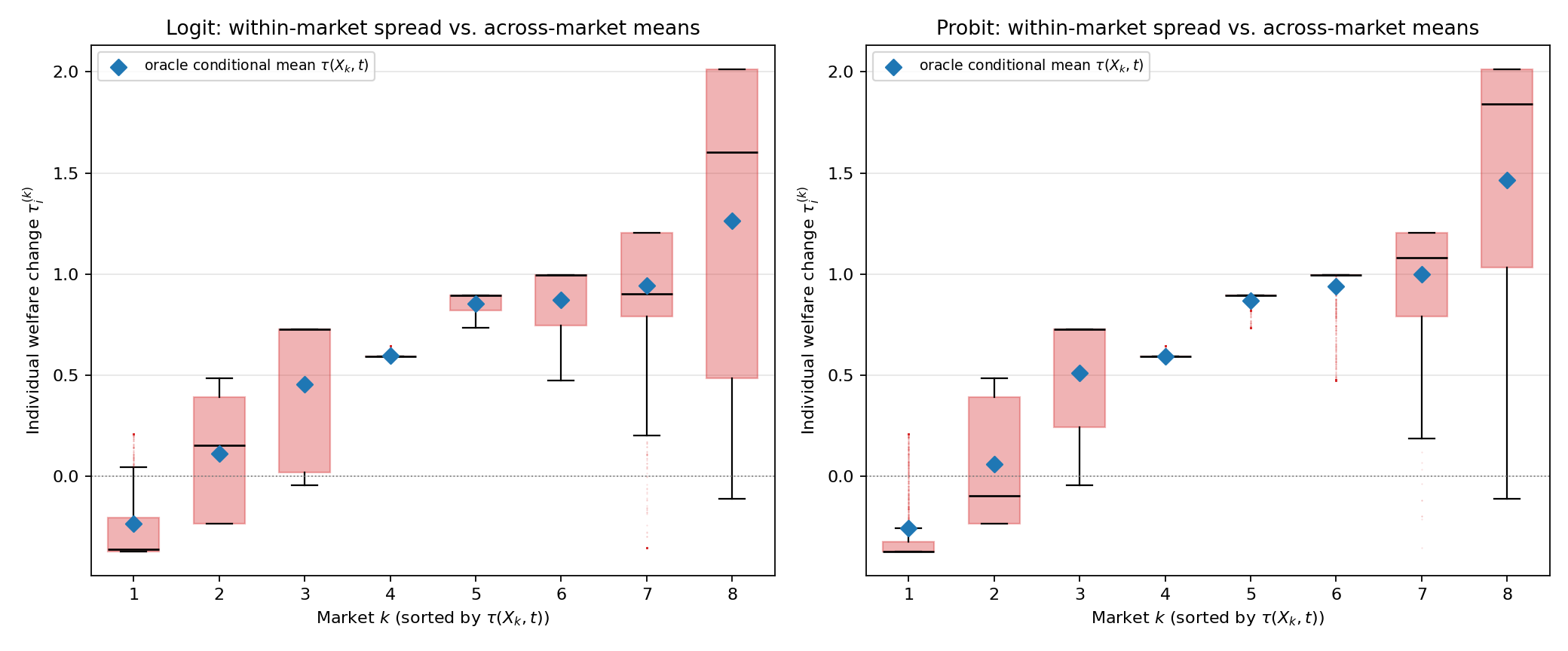}
		\caption{Within-market spread of simulated individual welfare changes $\tau_i^{(k)}$ (boxplots, sorted by $\tau(X_k,t)$) versus the oracle conditional means (diamonds). Markets vary substantially in within-market dispersion, which is the source of Theorem \ref{Bound_Change_Utility_Lower}'s looseness.}
		\label{fig:layer1_hetero}
	\end{figure}

\begin{figure}[ht]
		\centering
		\includegraphics[width=\textwidth]{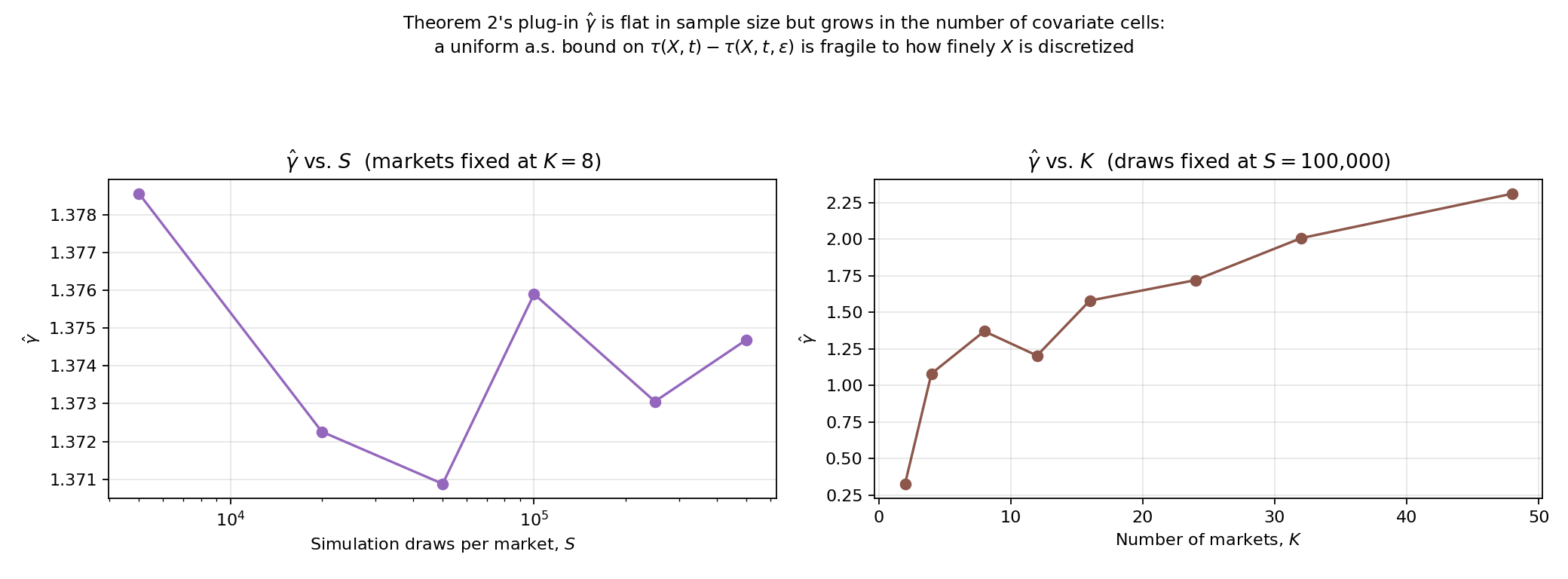}
		\caption{Sensitivity of the plug-in estimate $\hat\gamma$ to the number of simulation draws per market $S$ (left) and the number of covariate cells $K$ (right). $\hat\gamma$ is approximately flat in $S$ but grows steadily in $K$.}
		\label{fig:layer1_gamma}
	\end{figure}

	By contrast, Theorem \ref{Bound_Change_Utility_Lower}'s lower bound holds but is considerably looser, and the simulation reveals why. Figure \ref{fig:layer1_hetero} decomposes the pooled distribution by market, showing that within-market dispersion in $\tau_i^{(k)}$ varies substantially across markets: some markets exhibit almost no spread around their conditional mean, while others exhibit very wide spread. Because $\hat\gamma$ must bound the worst-case gap \emph{uniformly across all markets}, it is driven entirely by whichever single market happens to have the most extreme simulated individual, and Figure \ref{fig:layer1_gamma} shows that this plug-in estimate is essentially flat in the number of simulation draws $S$ once $S$ is moderately large, but grows substantially in the number of covariate cells $K$ (from $\hat\gamma\approx1.08$ at $K=4$ to $\hat\gamma\approx2.31$ at $K=48$ in our design). A uniform almost-sure bound on $\tau(X,t)-\tau(X,t,\varepsilon)$ therefore becomes \emph{less} informative, not more, as the analyst observes a finer partition of $X$, even though a finer partition intuitively conveys more information. This finding reinforces the case for the variance-based refinement of Theorem \ref{Bound_Change_Utility_Lower} suggested by a referee on an earlier version of this paper (paralleling Theorem 3.4 and Corollary 3.5 in \cite{Kallus2023}), which we view as a promising direction for future work.
	
	\subsection{Verifying Theorem \ref{Robust_Bound_Change_Utility}}\label{Sec:Sim_Layer2}
	
	We next illustrate the distributionally robust bound of Section \ref{Sec:Robust_Bounds}. Using the $K$ oracle conditional means $\tau(X_k,t)$ from the logit specification above as the empirical distribution $F_K$, we solve program (\ref{Robust_S_estimates}) under the $\chi^2$-divergence (Cressie--Read $k=2$, for which $\phi^*(s)=s^2/2+s$ is smooth and globally finite) across a grid of ambiguity radii $\rho\geq0$, for several values of $\beta$.
	
\begin{figure}[ht]
		\centering
		\includegraphics[width=0.75\textwidth]{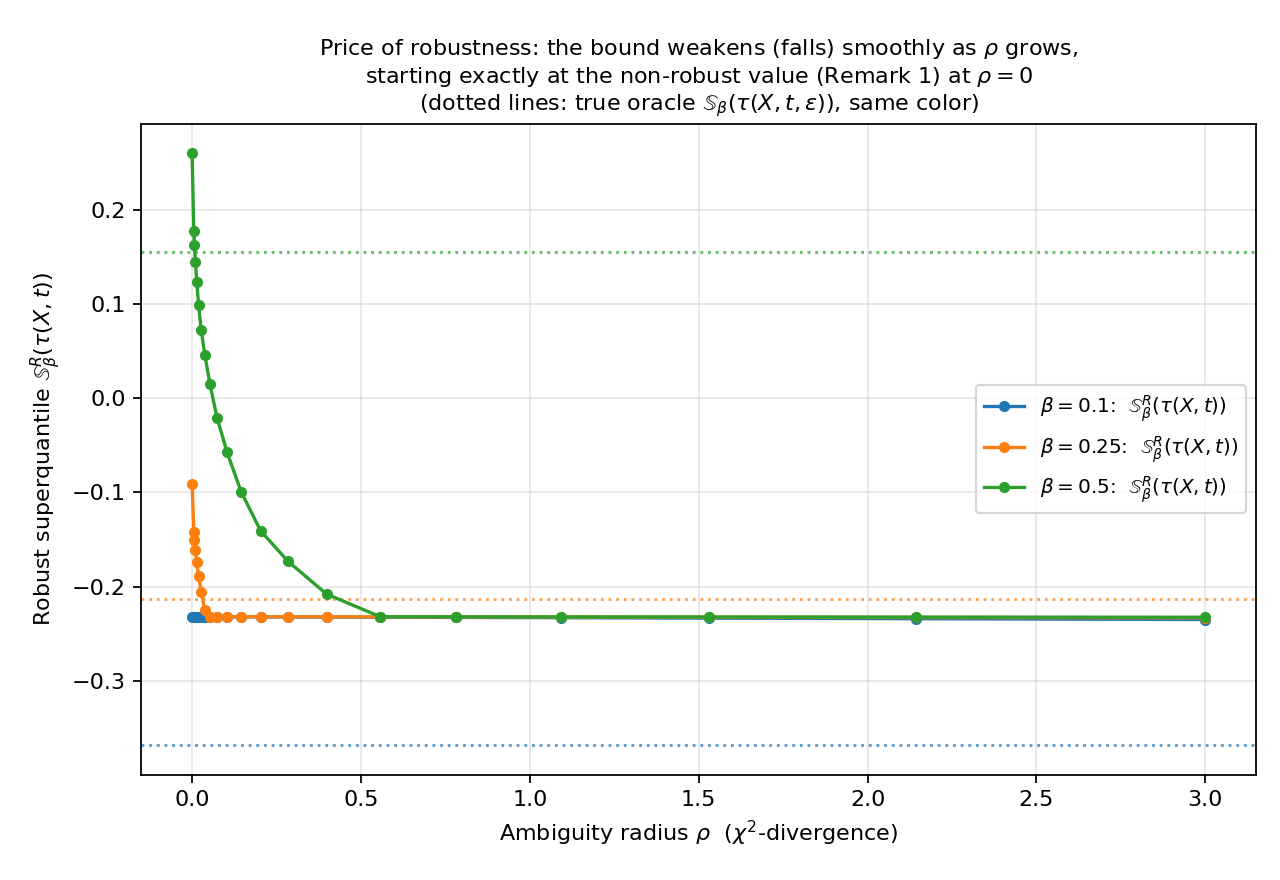}
		\caption{The robust upper bound $\mathbb{S}_\beta^R(\tau(X,t))$ as a function of the ambiguity radius $\rho$ under $\chi^2$-divergence, for $\beta\in\{0.1,0.25,0.5,0.9\}$. Each curve begins exactly at the corresponding non-robust value (Theorem \ref{Bound_Change_Utility}) at $\rho=0$ and weakens monotonically as $\rho$ grows. Dotted horizontal lines (same color) mark the true, unobserved oracle superquantile $\mathbb{S}_\beta(\tau(X,t,\varepsilon))$ for reference.}
		\label{fig:layer2_robustness}
	\end{figure}

	Figure \ref{fig:layer2_robustness} reports the resulting robust upper bound $\mathbb{S}_\beta^R(\tau(X,t))$ as a function of $\rho$. Two features of Theorem \ref{Robust_Bound_Change_Utility} are confirmed numerically. First, the $\rho=0$ nesting property of Remark \ref{Remark_Kallus_Nesting} holds exactly: the robust bound at $\rho=0$ coincides with the non-robust bound of Theorem \ref{Bound_Change_Utility} to within numerical tolerance, for every $\beta$ tested. Second, $\mathbb{S}_\beta^R(\tau(X,t))$ is non-increasing in $\rho$ across the entire grid, with zero violations: as the ambiguity set $\mathcal{U}_\phi(F,\rho)$ grows, the worst case over that set can only weaken the bound, never tighten it, since the infimum in (\ref{Robust_CVAR}) is taken over an enlarging feasible set. This monotonicity is precisely the ``price of robustness'' --- the cost, in terms of bound width, of insuring against ambiguity in the distribution of $X$ of a magnitude controlled by $\rho$.
	
	We now show that $\rho$ carries real, quantifiable value when the analyst's distribution of $X$ and the population to which the resulting welfare is applied do \emph{not} coincide. Fix the market weights $w^{\mathrm{train}}=F_K$ used above as a ``training'' population, and construct a genuinely different ``target'' population $w^{\mathrm{target}}$ by moving half the mass on the two highest-$\tau$ markets onto the two lowest-$\tau$ markets --- for instance, an eligibility expansion reaching a harder-hit population than the one on which the analyst's historical sample was collected. This shift has size $D_{\chi^2}(w^{\mathrm{target}}\|w^{\mathrm{train}})=0.080$. Rather than observing $w^{\mathrm{train}}$ exactly, the analyst estimates it from a sample of $N$ market-membership observations, and it is this finite-sample estimate $\hat w^{(N)}$, not $w^{\mathrm{train}}$ itself, that supplies $F_K$ for both the non-robust ($\rho=0$) and robust bounds. We say a bound $B$ \emph{covers} the true target value $T=\mathbb{S}_\beta^{w^{\mathrm{target}}}(\tau(X,t))$ when $B\leq T$: the reported number is then an appropriately conservative account of how badly off the worst-affected $\beta$-fraction of the deployment population actually is; $B>T$ is a false reassurance.
	
	Even absent estimation error --- using $w^{\mathrm{train}}$ itself as $F_K$ --- the non-robust bound fails outright: at $\beta=0.25$ it reports $-0.091$, against a true target value of $-0.206$, a failure driven entirely by the population shift. Theorem \ref{Robust_Bound_Change_Utility} closes this gap: $\mathbb{S}_\beta^R(w^{\mathrm{train}},\rho)$ falls monotonically in $\rho$ and crosses below the true target value at $\rho^\ast=0.028$, comfortably inside the divergence $D_{\chi^2}(w^{\mathrm{target}}\|w^{\mathrm{train}})=0.080$ that formally guarantees coverage --- the theorem's sufficient condition is conservative relative to what is numerically needed here.
	
\begin{figure}[ht]
		\centering
		\includegraphics[width=0.75\textwidth]{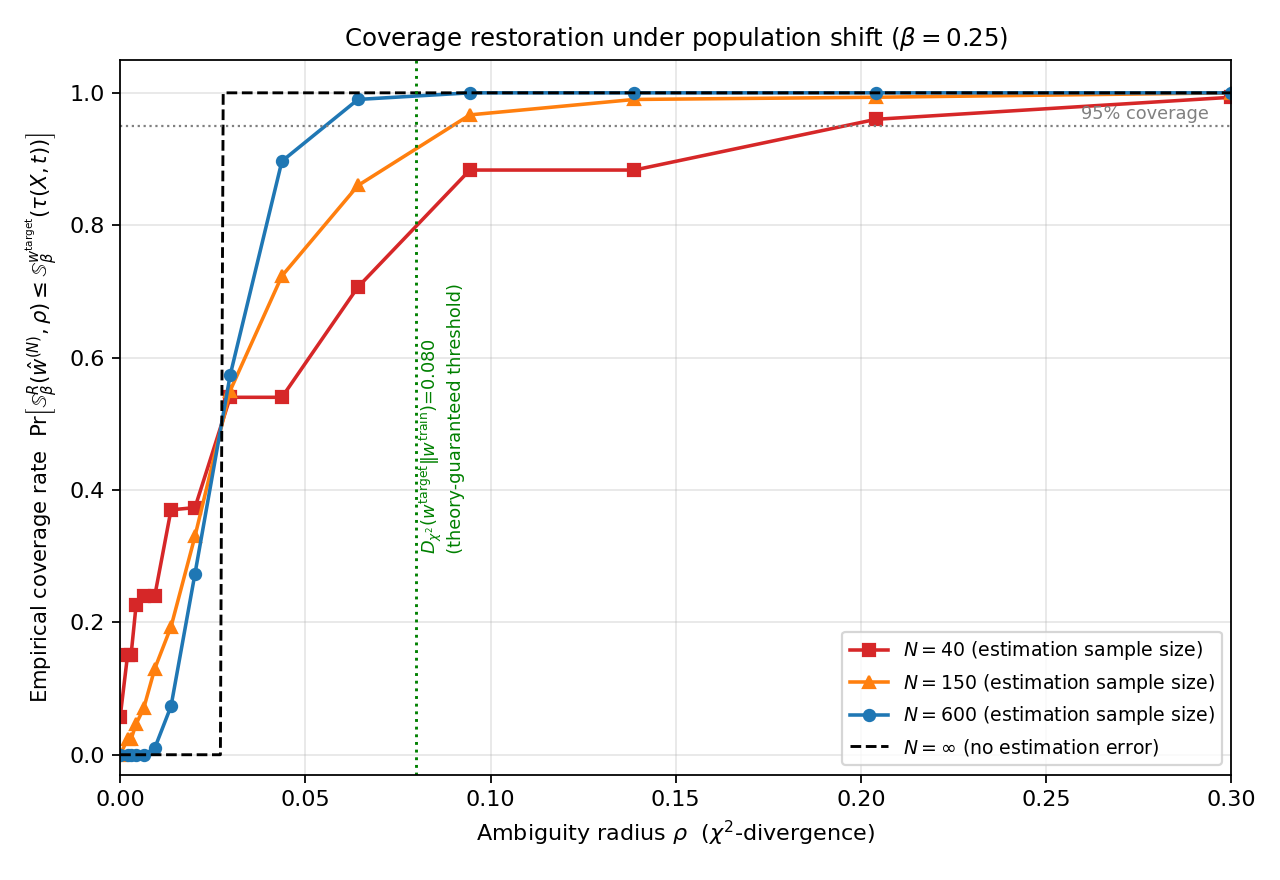}
		\caption{Empirical coverage rate of the robust bound $\mathbb{S}_\beta^R(\hat w^{(N)},\rho)$ against the true target value $\mathbb{S}_\beta^{w^{\mathrm{target}}}(\tau(X,t))$ ($\beta=0.25$), as a function of the ambiguity radius $\rho$, for estimation-sample sizes $N\in\{40,150,600\}$ (300 Monte Carlo replications each). The dashed step function is the deterministic $N=\infty$ benchmark, jumping at $\rho^\ast=0.028$; the dotted vertical line marks the divergence $D_{\chi^2}(w^{\mathrm{target}}\|w^{\mathrm{train}})=0.080$ that formally guarantees coverage.}
		\label{fig:layer2_coverage}
	\end{figure}

	Figure \ref{fig:layer2_coverage} adds finite-sample estimation error on top of the shift. For $N\in\{40,150,600\}$ estimation-sample sizes, each with 300 Monte Carlo draws of $\hat w^{(N)}$, it plots the empirical coverage rate $\Pr[\mathbb{S}_\beta^R(\hat w^{(N)},\rho)\leq \mathbb{S}_\beta^{w^{\mathrm{target}}}(\tau(X,t))]$ against $\rho$. Coverage is essentially zero at $\rho=0$ for every $N$: the plug-in bound is over-optimistic under this shift regardless of sample size. As $\rho$ grows, coverage rises smoothly toward one, crossing 95\% at $\rho\approx0.06$ for $N=600$, $\rho\approx0.09$ for $N=150$, and $\rho\approx0.20$ for $N=40$: a smaller estimation sample requires a larger $\rho$ to deliver the same guarantee, and all three curves converge toward the deterministic ($N=\infty$) threshold $\rho^\ast=0.028$ as $N$ grows. Choosing $\rho$ on the order of the anticipated population shift, with a sample-size-dependent margin for estimation error, thus delivers a numerically verified welfare guarantee that the non-robust bound does not.
	
	\section{Application}\label{Sec:Applicability_Distributional_WDZ}
	While the results in the preceding sections are general, this section illustrates how they apply to standard economic settings, beginning with treatment effects in a setting with endogenous participation and subjective treatment participation costs, where we focus on the benefits, costs, and welfare implications of social programs, and in particular on distributional results for the treated that are new to this literature. We then connect these results to the multi-outcome IV framework of \cite{Heckman_Urzua_Vytlacil_2006_REStat}, showing that our distributionally robust bounds remain valid across all instruments whose implied covariate distributions lie within the ambiguity set, at a price governed by an observable divergence between instrument-implied complier distributions that we quantify in Section \ref{Sec:Sim_Layer3}.
	
	\subsection{Treatment participation costs and welfare in generalized Roy models}\label{Sec:Gen_Roy_Cost} We begin with our most novel application: the welfare framework introduced by \cite{Eisenhauer_Heckman_Vytlacil_2015}, which, in the context of treatment effects, examines the marginal benefits and marginal costs of policies. Their framework extends the modern treatment-effect literature by providing a method to identify both the marginal benefits and the marginal costs of policy interventions. In particular, they incorporate agents' subjective costs associated with participation in social programs, allowing them to analyze the benefits, costs, and surplus (benefits minus costs) of specific social programs in average terms. We integrate their results with our framework to show how policymakers can assess not only the average benefits, costs, and surpluses of different policies, but also their distributional implications --- and, in Section \ref{Sec:TT}, the implications specifically for those who participate.
	
	\subsubsection{\cite{Eisenhauer_Heckman_Vytlacil_2015}'s framework}
	We focus on the traditional binary case. We assume that there are two potential outcomes $\left(V_0, V_1\right)$ and a choice indicator $D$, with $D=1$ if the agent selects into treatment so that $V_1$ is observed and $D=0$ if the agent does not select into treatment so that $V_0$ is observed. We use the potential outcome equation to denote the value, $V$, as $V=DV_1+(1-D)V_0$. Following \cite{Eisenhauer_Heckman_Vytlacil_2015},  we assume a separable structure in the outcomes  where $\EE\left[V_j \mid X\right]=\mu_j(X)$ and
	\begin{equation}\label{Binary_Outcome_add}
		V_j=\mu_j(X)+\nu_j\quad\mbox{for $j=0,1$}.
	\end{equation}

	As in previous sections,  $X$ is a (random) vector of covariates observed by the analyst, while ($\nu_0, \nu_1$) are unobserved (to the analyst) heterogeneity terms. Using (\ref{Binary_Outcome_add}), we can re-write $V$ as
	\begin{equation}\label{Switching_Outcomes_eq}
		V=\mu_0(X)+\left\{\left[\mu_1(X)-\mu_0(X)\right]+\nu_1-\nu_0\right\} D+\nu_0.
	\end{equation}
	
	Let $B=V_1-V_0$ denote the individual gross benefit of treatment, defined as the causal effect on $V$ of moving an otherwise identical individual from state 0 to state 1. Thus, $B$ measures the ceteris paribus change in the outcome induced by treatment.
	\smallskip
	
	Let $C$ denote the agent's \emph{subjective} treatment cost, defined as:
	\begin{equation}\label{Cost_C}
		C=\mu_C(Z)+\nu_C,   
	\end{equation}
	where  $Z$  represents an observed random vector of cost shifters and $\nu_C$ is a random variable unobserved by the analyst. Under this specification, the conditional expectation in  Eq. (\ref{Cost_C}) satisfies $\EE[C \mid Z] = \mu_C(Z)$.
	\smallskip
	
	Individuals choose to participate in the treatment if the perceived benefit from participation is greater than the subjective cost:
	\begin{equation}\label{D_choice}
		D= \begin{cases}1 & \text { if } \mathcal{W} \geq 0 \\ 0 & \text { otherwise }\end{cases}
	\end{equation}
	where $\mathcal{W}$  is the  \emph{individual welfare} (surplus), that is, the net benefit, from treatment:
	
	$$
	\begin{aligned}
		\mathcal{W} & \triangleq\left(V_1-V_0\right)-C \\
		& =\mu_W(X, Z)-\varepsilon_\mathcal{W} ,
	\end{aligned}
	$$
	with $\mu_\mathcal{W}(X, Z)=\left[\mu_1(X)-\mu_0(X)\right]-\mu_C(Z)$ and $\varepsilon_{\mathcal{W}}=\nu_C-\left(\nu_1-\nu_0\right)$.
	\smallskip

	Our distributional analysis of treatment costs and welfare does not impose functional-form restrictions on $\mu_0$, $\mu_1$, or $\mu_C$, nor does it require parametric assumptions on the distributions of $\nu_0$, $\nu_1$, or $\nu_C$. Instead, our results build on the identification framework of \cite{Eisenhauer_Heckman_Vytlacil_2015}, which enables the analyst to recover the relevant conditional average objects needed for our bounds.
	
	It is easy to see that the choice model (\ref{D_choice}) corresponds to a particular instance of the PUM. To formalize this, let  $ P(X, Z)\triangleq \operatorname{Pr}(D=1 \mid X$, $Z)$ denote the probability of selecting into treatment given $(X,Z)$. Given the structure of this cross-threshold model, we note that $P(X, Z)=F_{\varepsilon_{\mathcal{W}}}\left(\mu_\mathcal{W}(X, Z)\right)$, where $F_{\varepsilon_{\mathcal{W}}}(\cdot)$ denotes the distribution of $\varepsilon_{\mathcal{W}}$. For ease of exposition,  we denote $P(X, Z)$ by $P$, suppressing the $(X, Z)$ argument.  In addition, we use the fact that $U_\mathcal{W}=$ $F_{\varepsilon_\mathcal{W}}(\varepsilon_{\mathcal{W}})$ is a uniform random variable. In particular, different values of $\overline{\varepsilon}_\mathcal{W}$ denote different quantiles of $\varepsilon$. Under our previous assumptions, $F_{\varepsilon_\mathcal{W}}$ is strictly increasing, and $P(X, Z)$ is continuous conditional on $X$. Throughout this section, we assume the following.
	\begin{assumption}\label{Generalized_Roy_Assumpton}  The following conditions are assumed to hold:
		\begin{itemize}
			\item[(i)] $( \nu_0, \nu_1, \nu_C)$ is independent of $(X,Z)$.
			\item[(ii)]The distribution of $\mu_C(Z)$ conditional on $X$ is absolutely continuous with respect to Lebesgue measure.
			\item [(iii)] The distribution of $\varepsilon_{\mathcal{W}}=\nu_C-\left(\nu_1-\nu_0\right)$ is absolutely continuous with respect to Lebesgue measure and has a cumulative distribution function that is strictly increasing.
			\item[(iv)]  The population means $\EE[\left|V_1\right|], \EE[\left|V_0\right|]$, and $\EE[|C|]$ are finite.
		\end{itemize}

	\end{assumption}
	
	\subsubsection{Parameters}Our goal is to learn about the distributions of benefits, costs, and welfare---objects that are unobserved by the analyst. To do so, we rely on several key parameters introduced by \cite{Eisenhauer_Heckman_Vytlacil_2015}, which provide identification of the relevant conditional average objects that our analysis builds upon.
	\smallskip
	
	The first parameter is the conditional average treatment  effect (ATE) benefit given by:
	$$
	B^{\operatorname{ATE}}(x) \triangleq  \EE\left[V_1-V_0 \mid X=x\right]=\mu_1(x)-\mu_0(x) .
	$$

	$B^{\operatorname{ATE}}(x)$ denotes the ATE for individuals with characteristics $X = x$: that is, the causal effect of assigning treatment randomly to all individuals of type $x$, under full compliance and abstracting from general equilibrium or spillover effects.
	\smallskip

	The second parameter of interest is the average treatment benefit for individuals who actually receive the treatment, commonly referred to as the \emph{benefit of treatment on the treated}.
	$$\begin{aligned} B^{\mathrm{TT}}(x) & \triangleq \EE\left[V_1-V_0 \mid X=x, D=1\right]\\ & =\mu_1(x)-\mu_0(x)+\EE\left[\nu_1-\nu_0 \mid X=x, D=1\right].\end{aligned}$$

	\cite{Heckman_Vytlacil_1999,Heckman_Vytlacil_ECMA_2005_MTE} show that a uniform approach to $B^{\mathrm{ATE}}(x)$ and $B^{\mathrm{TT}}(x)$ is possible by using the MTE parameter, which is defined as:

	$$
	\begin{aligned}
		B^{\mathrm{MTE}}\left(x, u_\mathcal{W}\right) & \triangleq \EE\left[V_1-V_0 \mid X=x, U_\mathcal{W}=u_\mathcal{W}\right] \\
		& =\mu_1(x)-\mu_0(x)+\EE\left[\nu_1-\nu_0 \mid U_\mathcal{W}=u_\mathcal{W}\right] .
	\end{aligned}
	$$
	The function $B^{\mathrm{MTE}}\left(x, u_\mathcal{W}\right)$ is the treatment effect parameter that conditions the unobserved desire to select into treatment.
	\smallskip
	
	\cite{Eisenhauer_Heckman_Vytlacil_2015} note that conventional treatment-effects analysis does not define, identify, or estimate any component of treatment costs. To fill this gap, they introduce three cost parameters: the average cost of treatment, the average cost of treatment for those who select into treatment, and the marginal cost of treatment.

	$$
	\begin{aligned}
		C^{\mathrm{ATE}}(z) & =\EE[C \mid Z=z]=\mu_C(z), \\
		C^{\mathrm{TT}}(z) & = \EE[C \mid Z=z, D=1],\\ 
		&=\mu_C(z)+\EE\left[\nu_C \mid Z=z, D=1\right],\\
		C^{\mathrm{MTE}}\left(z, u_{\mathcal{W}}\right) & =\EE\left[C \mid Z=z, U_\mathcal{W}=u_\mathcal{W}\right], \\
		& =\mu_C(z)+\EE\left[\nu_C\mid U_\mathcal{W}=u_\mathcal{W}\right].
	\end{aligned}
	$$
	
	Finally, we define a set of welfare parameters. Recalling that  $\mathcal{W}=B-C=\mu_\mathcal{W}(X, Z)-\varepsilon_{\mathcal{W}}$ we get:
	$$\begin{aligned}
		\mathcal{W}^{\mathrm{ATE}}(x, z) & =\EE[\mathcal{W} \mid X=x, Z=z]=\mu_\mathcal{W}(x, z), \\
		\mathcal{W}^{\mathrm{TT}}(x, z) & =\EE[\mathcal{W} \mid X=x, Z=z, D=1]\\ & =\mu_\mathcal{W}(x, z)-\EE[\varepsilon \mid X=x, Z=z, D=1]
	\end{aligned}
	$$
	and
	$$\begin{aligned} 
		\mathcal{W}^{\mathrm{MTE}}\left(x, z, u_\mathcal{W}\right) & =\EE\left[\mathcal{W} \mid X=x, Z=z, U_\mathcal{W}=u_\mathcal{W}\right] \\
		& =\mu_\mathcal{W}(x, z)-\EE\left[\varepsilon_{\mathcal{W}} \mid U_\mathcal{W}=u_\mathcal{W}\right]\\
		& =\mu_\mathcal{W}(x, z)-F^{-1}_{\varepsilon_{\mathcal{W}}}\left(u_\mathcal{W}\right),\end{aligned}$$
	where the last equality uses $\varepsilon_{\mathcal{W}}=F^{-1}_{\varepsilon_{\mathcal{W}}}(U_{\mathcal{W}})$, which holds almost surely under Assumption \ref{Generalized_Roy_Assumpton}(iii). Equivalently, $\mathcal{W}^{\mathrm{MTE}}(x,z,u_{\mathcal{W}})$ equals $B^{\mathrm{MTE}}(x,u_{\mathcal{W}})-C^{\mathrm{MTE}}(z,u_{\mathcal{W}})$, and it vanishes at $u_{\mathcal{W}}=P(x,z)$: agents at the margin of participation are indifferent.
	
	\smallskip
	
	\cite{Eisenhauer_Heckman_Vytlacil_2015}  show how to identify the previous parameters. Proposition \ref{Bound_Welfare_MTE} below shows that their average identification results help us learn about the distributional aspects of welfare. 
	\begin{proposition}\label{Bound_Welfare_MTE} Let Assumption \ref{Generalized_Roy_Assumpton} hold. Then for $\beta\in(0,1]$
		
		\begin{equation}\label{Gen_Roy_B1}
			\mathbb{S}_\beta(\mathcal{W})\leq  \mathbb{S}_\beta(\mathcal{W}^{\operatorname{ATE}}(X,Z))
		\end{equation}
		and
		
		\begin{equation}\label{Gen_Roy_B2}
			\mathbb{S}_\beta(\mathcal{W})=  \mathbb{S}_\beta(\mathcal{W}^{\operatorname{MTE}}(X,Z, U_\mathcal{W})).
		\end{equation}
	\end{proposition}
	
	Proposition \ref{Bound_Welfare_MTE} shows how the two average welfare parameters in \cite{Eisenhauer_Heckman_Vytlacil_2015} reveal the unobserved distribution of $\mathcal{W}$. Intuitively, $\mathbb{S}_{\beta}(\mathcal{W})$ captures the welfare effect among the worst-off $(100\times\beta)\%$ of individuals, while $\mathbb{S}_{\beta}(\mathcal{W}^{\operatorname{ATE}}(X,Z))$ in bound (\ref{Gen_Roy_B1}) captures the worst outcomes only across groups defined by $(X,Z)$, and is therefore an upper bound. Conditioning additionally on $U_{\mathcal{W}}$ removes all remaining heterogeneity in welfare: because $\mathcal{W}=\mu_{\mathcal{W}}(X,Z)-F^{-1}_{\varepsilon_{\mathcal{W}}}(U_{\mathcal{W}})$ is a function of $(X,Z,U_{\mathcal{W}})$, we have $\mathcal{W}=\mathcal{W}^{\operatorname{MTE}}(X,Z,U_{\mathcal{W}})$ almost surely, and (\ref{Gen_Roy_B2}) holds with equality. Hence, whenever $\mathcal{W}^{\operatorname{MTE}}$ is identified over the relevant support (as in \cite{Eisenhauer_Heckman_Vytlacil_2015}), the entire distribution of individual welfare---and in particular $\mathbb{S}_\beta(\mathcal{W})$ for every $\beta$---is identified, not merely bounded. The gap between (\ref{Gen_Roy_B1}) and (\ref{Gen_Roy_B2}) thus measures exactly the role of unobserved heterogeneity in the distribution of welfare. This is special to welfare: the gross benefit $B$ and the cost $C$ are not degenerate given $(X,Z,U_{\mathcal{W}})$, which is why the results for costs below remain genuine bounds. To the best of our knowledge, Proposition \ref{Bound_Welfare_MTE} is new in the context of generalized Roy models with participation costs.
	\smallskip
	
	Note that (\ref{Gen_Roy_B1}) and (\ref{Gen_Roy_B2}) explicitly incorporate the cost of treatment participation. Consequently, the result provides information about potential welfare losses across subpopulations as a function of observables $(X,Z)$, the cost $C$, and unobserved heterogeneity $U_{\mathcal{W}}$. Furthermore, our framework extends \cite{Eisenhauer_Heckman_Vytlacil_2015}'s results by providing lower bounds that identify which groups incur the highest costs. To do so, we focus on the right superquantile $\overline{\mathbb{S}}_\alpha(\cdot)$, which---as discussed earlier---captures the average outcome in the upper tail of the distribution and is therefore the appropriate object for assessing individuals who face the highest participation costs.

	\begin{proposition}\label{C_lower_bound}Let Assumption \ref{Generalized_Roy_Assumpton} hold.  Then
		\begin{equation}\label{Lower_bound_cost_ATE}
			\overline{\mathbb{S}}_\alpha(C)\geq \overline{\mathbb{S}}_\alpha(C^{\operatorname{ATE}}(Z))\quad \text{for $\alpha\in[0,1)$}
		\end{equation}
		and 
		\begin{equation}\label{Lower_bound_cost_MTE}
			\overline{\mathbb{S}}_\alpha(C)\geq \overline{\mathbb{S}}_\alpha(C^{\operatorname{MTE}}(Z,U_\mathcal{W}))
		\end{equation}  
	\end{proposition}

	The previous result provides information on the ATE and MTE costs of the $(100\times (1-\alpha ))\%$ worst affected. Expression (\ref{Lower_bound_cost_ATE}) provides a bound in terms of observables $Z$. Intuitively, this lower bound lets the analyst identify which subgroups bear a higher cost. Accordingly, this bound can guide the reduction of treatment costs 
	in target subgroups, which can be welfare-improving. Similarly, bound (\ref{Lower_bound_cost_MTE}) complements the previous analysis by incorporating the role of unobservables.
	
	Finally, we point out that \cite{Eisenhauer_Heckman_Vytlacil_2015}'s analysis is conducted in terms of average costs. Proposition \ref{C_lower_bound} shows that their results inform the distributional implications of treatment costs, accounting for both observables and unobservables. 
	
	\subsection{Treatment on the treated}\label{Sec:TT} We now turn to our most novel result: the distributional properties of $\mathcal{W}^{\operatorname{TT}}$ and $C^{\operatorname{TT}}$.\footnote{For ease of exposition, we omit the analysis of $B^{\operatorname{TT}}$. However, all arguments extend directly to that case.} Because of the missing-outcome problem, neither distribution is observable to the analyst (\cite{HECKMAN2007_Handbook_partI}). Our objective is to exploit the identification results in \cite{Eisenhauer_Heckman_Vytlacil_2015} for the parameters $\mathcal{W}^{\operatorname{TT}}(X,Z)$ and $C^{\operatorname{TT}}(Z)$ in order to recover informative bounds on the unknown distributions of $\mathcal{W}^{\operatorname{TT}}$ and $C^{\operatorname{TT}}$, conditioning on realized treatment status $D=1$ rather than on a policy assignment as in Section \ref{Sec:Gen_Roy_Cost}.
	\smallskip

	In doing so,  we make use of the following notation:
	\begin{eqnarray*}
		\mathbb{S}_\beta(\mathcal{W}^{\operatorname{TT}})&\triangleq&\mathbb{S}_\beta(\mathcal{W}\mid D=1),\\
		&=&\sup_{\lambda \in \RR}\left\{\lambda+{1\over\beta}\EE[(\mathcal{W}-\lambda)_{-}\mid D=1]\right\}.
	\end{eqnarray*}
	
	In a similar way, for the cost $C$ we define :
	\begin{eqnarray*}
		\overline{\mathbb{S}}_\alpha(C^{\operatorname{TT}})&\triangleq&\mathbb{S}_\alpha(C\mid D=1),\\
		&=&\inf_{\lambda\in\RR}\left\{\lambda+{1\over 1-\alpha}\EE[(C-\lambda)_{+}\mid D=1]\right\}.
	\end{eqnarray*}

	Intuitively,
	$\mathbb{S}_\beta\!\left(\mathcal{W}^{\mathrm{TT}}\right)$
	is the superquantile of the welfare distribution for the treated. It equals the average welfare among the $(100\times \beta)\%$ of treated individuals with the lowest welfare gains, including welfare losses. Similarly, $\overline{\mathbb{S}}_\alpha(C^{\mathrm{TT}})$
	is the superquantile of the treatment-cost distribution for the treated. It corresponds to the average treatment cost for the $(100\times (1-\alpha))\%$ of treated individuals who incur the highest program costs.
	\smallskip
	
	The following is the main result of this section.

	\begin{theorem}\label{Wel_Cost_TT_bounds}Let Assumption \ref{Generalized_Roy_Assumpton} hold. Then:
		\begin{itemize}
			\item[(i)]For $\beta\in(0,1]$
			\begin{equation}\label{Welfare_TT_Bound}
				\mathbb{S}_\beta(\mathcal{W}^{TT})\leq \mathbb{S}_\beta(\mathcal{W}^{TT}(X,Z)).
			\end{equation}
			\item[(ii)] For $\alpha\in[0,1)$
			\begin{equation}\label{Cost_TT_Bound}
				\overline{\mathbb{S}}_\alpha(C^{\operatorname{TT}})\geq \overline{\mathbb{S}}_\alpha(C^{\operatorname{TT}}(X,Z))
			\end{equation}
		\end{itemize}
	\end{theorem}

	Some remarks are in order. First, part (i) bounds the welfare of the $(100\times \beta)\%$ worst-affected individuals in the population. To see the relevance of bound (\ref{Welfare_TT_Bound}), assume without loss of generality that
	$\mathcal{W}$ has a continuous distribution. In this case,
	$\mathbb{S}_\beta(\mathcal{W}^{\mathrm{TT}})
	= \mathbb{E}\!\left[\mathcal{W}\mid \mathcal{W}\le F_{\mathcal{W}}^{-1}(\beta),\, D=1\right]$.
	When only the cell-level parameter $\mathcal{W}^{\mathrm{TT}}(X,Z)$ is available---for instance, with discrete instruments or limited support of the propensity score---neither the distribution of $\mathcal{W}$ nor its conditional distribution given $D=1$ is identified.\footnote{Under full support, by contrast, $\mathcal{W}=\mu_{\mathcal{W}}(X,Z)-F^{-1}_{\varepsilon_{\mathcal{W}}}(U_{\mathcal{W}})$ is a function of $(X,Z,U_{\mathcal{W}})$, and its distribution among the treated, $\{D=1\}=\{U_{\mathcal{W}}\le P(X,Z)\}$, is identified whenever $\mathcal{W}^{\mathrm{MTE}}$ is; see the discussion following Proposition~\ref{Bound_Welfare_MTE}. Bound (\ref{Welfare_TT_Bound}) is therefore most useful when only cell-level parameters are available, whereas the cost bound in part (ii) remains informative in both cases, because $C$ is not a function of $(X,Z,U_{\mathcal{W}})$.} Thus
	$\mathbb{E}\!\left[\mathcal{W}\mid \mathcal{W}\le F_{\mathcal{W}}^{-1}(\beta), D=1\right]$ is not observed by the analyst. By contrast, using the identified function $\mathcal{W}^{\mathrm{TT}}(X,Z)$, we can construct the upper bound
	$\mathbb{E}\!\left[\mathcal{W}(X,Z)\mid \mathcal{W}(X,Z)\le F_{\mathcal{W}(X,Z)}^{-1}(\beta),\, D=1\right]$,
	which provides an informative measure of the potential gains or losses experienced by the bottom $(100\times \beta)\%$ of the welfare distribution among treated individuals. Thus, based on observables, the policymaker can identify which treated subgroups are likely to experience declines in welfare.
	\smallskip
	
	Part (ii) establishes a lower bound on the treatment cost among individuals who received the treatment. As with $\mathcal{W}^{\mathrm{TT}}$, the conditional distribution of $C^{\mathrm{TT}}$ is not identified. However, using the results in \cite{Eisenhauer_Heckman_Vytlacil_2015}, we can identify the lower bound (\ref{Cost_TT_Bound}) and compute it using the linear programming procedure described in Section \ref{sec{computin_LP}}. From an economic perspective, the lower bound (\ref{Cost_TT_Bound}) is informative along at least two dimensions. First, it enables the analyst to determine which subpopulations bear higher treatment costs among the treated. Second, it sheds light on the fairness of treatment costs across groups---for example, whether a particular intervention is regressive or progressive.  
	\smallskip
	
	Theorem \ref{Wel_Cost_TT_bounds}(i) also admits the same probabilistic reading as Proposition \ref{Prop:bPOE}, now specialized to program participants.
	
	\begin{corollary}[Buffered probability of welfare loss among the treated]\label{Cor:bPOE_TT}
		Let Assumption \ref{Generalized_Roy_Assumpton} hold, and fix a threshold $\theta$ in the common domain of $p_\theta(\mathcal{W}^{TT})$ and $p_\theta(\mathcal{W}^{TT}(X,Z))$. Then
		\begin{equation}\label{bPOE_TT_bound}
			p_\theta(\mathcal{W}^{TT}(X,Z))\;\leq\;p_\theta(\mathcal{W}^{TT}).
		\end{equation}
	\end{corollary}
	
	For a policy-relevant loss threshold $\theta$---a specific dollar amount of forgone welfare, say---$p_\theta(\mathcal{W}^{TT}(X,Z))$, computable from the identified function $\mathcal{W}^{TT}(X,Z)$ alone, is a valid lower bound on the true buffered probability that a treated individual experiences a welfare loss no better than $\theta$. This is a more directly actionable statement than the magnitude bound in Theorem \ref{Wel_Cost_TT_bounds}(i) alone: a program administrator setting a compensation threshold, a hardship exemption, or an eligibility rule for supplemental assistance can ask directly what share of participants are estimated to fall below it, without first fixing a tail fraction $\beta$ or observing individual-level outcomes.
	\smallskip
	
	We close this section by noting that, to the best of our knowledge, Theorem \ref{Wel_Cost_TT_bounds} and Corollary \ref{Cor:bPOE_TT} are new in the cost-benefit analysis of treatment effects.
	\smallskip
	
	The bounds in Theorem~\ref{Wel_Cost_TT_bounds} take the distribution of observable characteristics $X$ among the treated as given. In practice, however, the analyst may face additional uncertainty about this distribution: the composition of program participants can shift across cohorts, sites, or policy regimes, so that the empirical distribution $F_K$ estimated from one sample may not accurately reflect the covariate distribution of the population to which the bounds will be applied. In such settings, the distributionally robust extension of Section~\ref{Sec:Robust_Bounds} applies directly. Replacing $F_K$ with the ambiguity set $\mathcal{U}_\phi(F_K,\rho)$ in program~(\ref{Robust_S_estimates}), the robust treatment-on-the-treated welfare bound
	\begin{equation}\label{eq:robust_WTT}
		\mathbb{S}^{R}_\beta\!\left(\mathcal{W}^{\operatorname{TT}}(X,Z)\right)
		\;=\;
		\inf_{G\,\in\,\mathcal{U}_\phi(F_K,\rho)}\;
		\mathbb{S}^{G}_\beta\!\left(\mathcal{W}^{\operatorname{TT}}(X,Z)\right)
	\end{equation}
	remains valid for any covariate distribution within $\rho$ of $F_K$ in $\phi$-divergence, by the same argument as Theorem~\ref{Robust_Bound_Change_Utility}. The radius $\rho$ thus quantifies how sensitive the welfare-at-risk conclusion is to uncertainty about who the treated are --- a question of direct policy relevance when program eligibility or take-up rates are expected to change.
	\subsection{Connection to multi-outcome IV: instrument sensitivity and robust welfare risk}\label{Sec:HUV_Connection}
	
	The results of the preceding sections deliver distributional welfare bounds for the binary generalized Roy model. We now situate these results within the broader multi-outcome framework of \cite{Heckman_Urzua_Vytlacil_2006_REStat} (hereafter HUV), which provides a natural motivation for why robustness is indispensable in welfare analysis beyond the binary case.
	\smallskip
	
	HUV extends the MTE framework to settings with more than two outcomes, covering both ordered and unordered choice models. A central finding of their Section~VI is that in ordered choice settings, the standard IV estimand is a weighted average of transition-specific MTEs,
	\begin{equation}\label{eq:transition_mte}
		\Delta^{\mathrm{MTE}}_{s,s+1}(u_{\mathcal{W}})
		\;=\;
		\EE\!\left(Y_{s+1} - Y_s \mid X = x,\, U_{\mathcal{W}} = u_{\mathcal{W}}\right),
	\end{equation}
	where the weights depend on the joint distribution of the instrument $J(Z,W)$ and the transition-specific propensity scores $\pi_s(Z,W_s)$. Crucially, HUV show (their Figures~10--11) that these weights need not be non-negative: even when every $\Delta^{\mathrm{MTE}}_{s,s+1}(u_{\mathcal{W}}) > 0$ pointwise, the IV estimand can be negative, and its sign and magnitude oscillate with the distribution of the instrument. Point-identified IV-based welfare estimates therefore inherit this instrument sensitivity in a fundamental way.
	\smallskip
	
	Our framework addresses precisely this fragility. Rather than targeting a point-identified weighted average of transition-specific MTEs, we work with the welfare functional
	\begin{equation}\label{eq:war_chain_huv}
		\mathbb{S}^{R}_\beta(\mathcal{W}) \;\leq\; \mathbb{S}^{R}_\beta(\mathcal{W}^{\mathrm{ATE}}(X,Z)) \;\leq\; \mathbb{S}_\beta(\mathcal{W}^{\mathrm{ATE}}(X,Z)),
	\end{equation}
	which integrates over the MTE distribution in a distributional sense rather than recovering a single weighted average. The key observation is that the transition-specific MTE \eqref{eq:transition_mte} is the building block of the welfare gain from moving an agent between sectors $s$ and $s+1$ in the generalized Roy model. Our welfare-at-risk functional $\mathbb{S}_\beta(\mathcal{W}^{\mathrm{ATE}}(X,Z))$ aggregates these gains across the distribution of $U_{\mathcal{W}}$, and our distributionally robust extension (Theorems~\ref{Robust_Bound_Change_Utility} and \ref{Robust_Bound_Change_Utility_Lower}) bounds the welfare superquantile over an ambiguity set $\mathcal{U}_\phi(F,\rho)$ that accommodates precisely the instrument-induced uncertainty that HUV documents.
	\smallskip
	
	To make this connection precise, consider the ordered Roy economy with $\bar{S}$ sectors and treatment-on-the-treated welfare cost
	\begin{equation}\label{eq:CTT_ordered_huv}
		\mathbb{S}_\beta(\mathcal{W}^{\mathrm{TT}})
		\;=\;
		\mathbb{S}_\beta\!\left(\mathcal{W} \mid D = 1\right),
	\end{equation}
	as defined in Section~\ref{Sec:TT}. In the binary Roy model, this is exactly the object bounded by Theorem~\ref{Wel_Cost_TT_bounds}. In the ordered multi-sector case, $\mathbb{S}_\beta(\mathcal{W}^{\mathrm{TT}})$ inherits the instrument sensitivity of $\Delta^{\mathrm{MTE}}_{s,s+1}$ when estimated via IV, since the weights on the transition-specific MTEs need not be non-negative (HUV, equations (32)--(33)). Our DRO bounds remain valid in this setting because they do not require the IV weights to be non-negative. Instead, the robust welfare risk $\mathbb{S}^{R}_\beta$ is defined as an infimum over $\mathcal{U}_\phi(F,\rho)$:
	\begin{equation}\label{eq:robust_war_huv}
		\mathbb{S}^{R}_\beta\!\left(\mathcal{W}^{\mathrm{ATE}}(X,Z)\right)
		\;=\;
		\inf_{G \in \mathcal{U}_\phi(F,\rho)}\;
		\mathbb{S}^{G}_\beta\!\left(\mathcal{W}^{\mathrm{ATE}}(X,Z)\right).
	\end{equation}
	Because the infimum is taken over the ambiguity set rather than over instrument choices, \eqref{eq:robust_war_huv} does not depend on which instrument the analyst uses to estimate $\pi_s(Z,W_s)$, provided the covariate distribution implied by that instrument lies in $\mathcal{U}_\phi(F,\rho)$: for every such instrument, the DRO bound \eqref{eq:robust_war_huv} holds uniformly. Section~\ref{Sec:Sim_Layer3} shows that the radius this requires, $\rho^{*}=D_\phi(F_2\Vert F_1)$, is computable from estimated propensity scores.
	\smallskip
	
	In the unordered case of HUV Section~VIB, where the multinomial choice problem reduces to a binary comparison of option $j$ versus the next-best alternative $\mathcal{J}\setminus j$, an analogous argument applies. The welfare cost
	\begin{equation}\label{eq:CTT_unordered_huv}
		\mathbb{S}_\beta(\mathcal{W}^{\mathrm{TT}}_j)
		\;=\;
		\mathbb{S}_\beta\!\left(Y_j - Y_{\mathcal{J}\setminus j} \mid I_{\mathcal{J}} = j\right)
	\end{equation}
	is the natural unordered-choice generalization where $I_{\mathcal{J}} =j$ denotes that alternative $j$ is chosen from set $\mathcal{J}$, and $\mathbb{S}^{R}_\beta$ bounds it robustly under the same $\mathcal{U}_\phi(F,\rho)$-ambiguity. This connection suggests that the framework developed here extends naturally to the richer multi-sector environments studied by HUV, providing distributional welfare guarantees that are unavailable from standard IV methods alone.
	\subsection{Instrument sensitivity and the price of immunity}\label{Sec:Sim_Layer3}
	
	Our final simulation illustrates the connection, developed formally in Section \ref{Sec:HUV_Connection}, between the robust bounds of Theorems \ref{Robust_Bound_Change_Utility}--\ref{Robust_Bound_Change_Utility_Lower} and the instrument-sensitivity problem of \cite{Heckman_Urzua_Vytlacil_2006_REStat}. The design is a reduced-form counterpart of the phenomenon they document, in the spirit of their Section VI: we consider a three-alternative unordered choice model over $K=12$ covariate cells, with logit unobserved heterogeneity and two binary instruments $Z_1$ and $Z_2$ that shift the utility of alternative $1$ with cell-specific strengths of opposite gradient across cells, indexed by a contrast parameter $c\geq0$. Because both instruments shift the same utility, per-complier outcome effects are identical across instruments within a cell; all instrument sensitivity therefore flows through the cell weighting that each instrument induces, exactly the channel underlying the negative-weights results of \cite{Heckman_Urzua_Vytlacil_2006_REStat}. Outcome levels are $Y_0=10$, $Y_1=15$, and $Y_2\in\{12,22\}$ across cells, with $Y_2=22$ specifically in the cells where alternative $2$ predominates. Because the subsidy raises the \emph{utility}, not the outcome, of alternative $1$, it can induce switches out of alternative $2$ even where $Y_2$ exceeds $Y_1$---a utility gain paired with an outcome loss---which is the source of the sign instability documented next.
	
	Two findings emerge. First, at contrast $c=0.25$ the two Wald estimands are $+1.60$ under $Z_1$ and $-0.43$ under $Z_2$: the estimands disagree even in sign, despite both instruments being valid, replicating in reduced form the sensitivity documented by \cite{Heckman_Urzua_Vytlacil_2006_REStat}. Second, letting $F_1$ and $F_2$ denote the complier-weighted covariate distributions implied by each instrument, the robust bound computed from nominal distribution $F_1$ at radius $\rho^{*}=D_{\phi}(F_2\Vert F_1)=0.093$ (Cressie--Read, $k=2$) equals $0.0930$ at $\beta=0.2$: it lies weakly below both instrument-specific non-robust bounds ($0.1081$ under $F_1$ and $0.0992$ under $F_2$), so it is a valid bound regardless of which instrument the analyst uses, while remaining strictly above the trivial floor $\min_x\tau(x)=0.0892$. Instrument-immunity is thus achieved at an informative level even as the point estimands contradict one another.
	
\begin{figure}[ht]
		\centering
		\includegraphics[width=\textwidth]{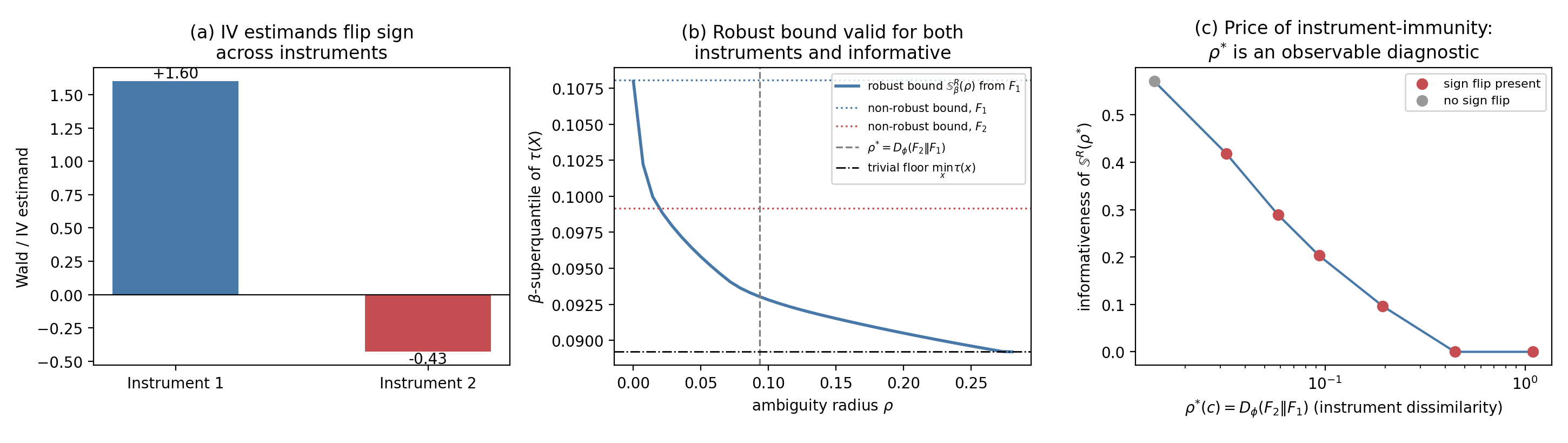}
		\caption{Instrument sensitivity and the price of immunity ($\beta=0.2$, Cressie--Read $k=2$). Panel (a): Wald estimands under the two instruments at contrast $c=0.25$ disagree in sign. Panel (b): the robust bound $\mathbb{S}^R_\beta(\rho)$ computed from $F_1$ crosses below the $F_2$ non-robust value before $\rho^{*}=D_\phi(F_2\Vert F_1)$ (dashed vertical line) and remains above the trivial floor $\min_x\tau(x)$ (dash-dotted line). Panel (c): informativeness of the robust bound at $\rho=\rho^{*}(c)$ against $\rho^{*}(c)$ as the instrument contrast varies; red markers indicate designs in which the estimand sign flip is present. The divergence $\rho^{*}$ is computable from estimated propensity scores and serves as a diagnostic for the cost of instrument-robustness.}
		\label{fig:layer3_immunity}
	\end{figure}

	The immunity is not free, and its price is governed by a quantity the analyst can observe. Sweeping the contrast $c$ traces out a frontier: as the instruments' complier distributions diverge, $\rho^{*}(c)$ grows and the informativeness of the robust bound at $\rho=\rho^{*}$ falls, where informativeness is defined as
	\begin{equation}\label{Informativeness_def}
		\operatorname{Inf}(\rho^{*})\;\triangleq\;\frac{\mathbb{S}_\beta^R(\rho^{*})-\min_x\tau(x)}{\mathbb{S}_\beta(\tau(X))\big|_{F_1}-\min_x\tau(x)},
	\end{equation}
	the robust bound's distance above the trivial floor relative to the non-robust bound. At the smallest contrast in this design, informativeness is roughly $57\%$ (at $\rho^{*}=0.014$); it falls to $42\%$ at $\rho^{*}=0.032$---the smallest contrast at which the sign flip appears in this design---and reaches zero by $\rho^{*}\approx0.45$. Figure \ref{fig:layer3_immunity} reports the estimands, the robust bound as a function of $\rho$, and this frontier. The practical implication is a diagnostic: since $\rho^{*}$ is a $\phi$-divergence between complier-weighted covariate distributions, it is computable directly from estimated propensity scores, so the analyst can assess \emph{before} robustifying whether instrument-immunity comes at an acceptable cost in the application at hand. Two remarks on scope. The design collapses the transition-specific MTE structure of \cite{Heckman_Urzua_Vytlacil_2006_REStat} into complier reweighting across cells; it is an illustration of the mechanism rather than a replication of their decomposition. And the ambiguity set is centered at $F_1$; since $\phi$-divergences are directional, centering at $F_2$ (or at the population distribution) yields a different required radius, and in applications the choice of nominal distribution should reflect which instrument-implied population the analyst regards as primary.
	
	\section{Conclusion}\label{Sec:Conclusion}
	This paper develops a distributional framework for analyzing welfare heterogeneity across individuals. By combining the concept of superquantiles with observable-group average welfare effects, we show how to construct informative bounds on individual welfare changes without relying on strong structural assumptions or detailed micro-level data.
	
	Although the framework is broadly applicable, much of the analysis is conducted within the class of perturbed utility models (PUMs), which offer an empirically tractable setting for nonparametrically estimating average welfare effects using standard data. We illustrate the usefulness of the framework with an application to social programs modeled through generalized Roy models with subjective participation costs, where we provide distributional bounds for the treated (treatment on the treated) that are new to this literature. We then connect these results to the multi-outcome IV framework of \cite{Heckman_Urzua_Vytlacil_2006_REStat}, showing that our distributionally robust bounds provide welfare guarantees that hold for all instruments whose implied covariate distributions lie within the ambiguity set, addressing the instrument-sensitivity problem they document. The core results established in Section \ref{Sec:Bounds_Ind_Welfare} extend naturally to this setting, even though the relevant welfare concepts and identification strategies differ.
	\bibliographystyle{plainnat}
	\bibliography{Quantile_References}
	
	\newpage
	\appendix
	
	\section{$\phi$-divergences}\label{App_phi-div}
	
	The following proposition summarizes the key properties of the class of  $\phi$-divergences we used in the paper.
	\begin{proposition}\label{phi_properties}The $\phi$-divergence functional (\ref{phi_divergence_definition}) is well-defined and nonnegative. It is equal to zero if and only if $f(t)=g(t)$ a.e. Furthermore, $D_{\phi}$ is convex on each of its arguments.
	\end{proposition}
	\proof The proof that $D_{\phi}(G\| F)$  is well defined and nonnegative follows from Proposition 1 in \cite{BentalTeboulle1987}.  The convexity of $D_{\phi}$ follows from Proposition 2 in \cite{BentalTeboulle1987}.\eproof
	
	In the main text  we define the  convex conjugate of $\phi(t)$ as $\phi^\ast(s)=\sup _{t \in \mathbb{R}}\{s t-\phi(t)\}$. From 
	Corollary 12.2.2 in \cite{Rockafellar1970} it follows that:
	\begin{equation}\label{Phi_convex_conjugate}
		\phi^*(s)=\sup _{t \in \mathbb{R}}\{s t-\phi(t)\}=\sup _{t \in \operatorname{dom} \phi}\{s t-\phi(t)\}=\sup _{t \in \operatorname{int} \operatorname{dom} \phi}\{s t-\phi(t)\}, 	\nonumber
	\end{equation}
	
	It is well-known that the conjugate $\phi^*$ is a closed proper convex function (\cite{Rockafellar1970}), with int dom $\phi^*=(a, b)$, where
	$$
	a=\lim _{t \rightarrow-\infty} t^{-1} \phi(t) \in[-\infty,+\infty) ; b=\lim _{t \rightarrow+\infty} t^{-1} \phi(t) \in(-\infty,+\infty] .
	$$
	Moreover, since $\phi$ is convex and closed, we have for its bi-conjugate $\phi^{* *}=\phi$, (\cite{Rockafellar1970}).
	It is worth noting that using the fact that  $1$ is the minimizer of $\phi$ and it is in the interior of its domain, so $\phi'(1) = 0$ holds. In addition,  using the property that $\phi$ is convex and closed,  we have by Fenchel equality 
	$ y =\phi'(x)$ iff $x = \phi^{*'}(y)$. Applying this latter observation to $x=1$ and $y=0$ we obtain $\phi^{\ast\prime}(0)=1$.
	\smallskip
	
	The following table Table \ref{tb:div} overviews three popular $\phi$-divergences satisfying Assumption

	\begin{table}[h]
		\begin{center}
			\begin{tabular}{ |c | c | c |c | c | c |}
				\hline
				Divergence & $\phi(t)$  &  $\phi^*(s)$ & Domain & $\phi^{*'}$  &$\phi^{*''}$ \\ \hline
				{\it Kullback-Leibler} & $t \log t$ & $e^{s-1}$ & $\RR$ & $e^{s-1}$ & $e^{s-1}$ \\ \hline
				{\it  Reverse Kullback-Leibler} & $- \log(t)$ & $ -1-\log(-s)$ & $\RR_{--}$ & $-\frac{1}{s}$  & $\frac{1}{s^2}$ \\ \hline
				{\it Hellinger Distance} & $(\sqrt{t}-1)^2$ & $\frac{s}{1-s}$ & $s < 1$ & $ \frac{1}{\left(1-s\right)^2}$ &  $-\frac{2}{\left(s-1\right)^3}$ \\ \hline
				
			\end{tabular}
			\caption{$\phi$-divergences with their convex conjugates and first and second derivatives.}\label{tb:div}
		\end{center}
	\end{table} 
	
	\newpage
	\section{Inference for the plug-in linear program}\label{App:LP_Inference}
	
	Throughout, $\beta\in(0,1)$, the weights $w_k>0$ with $\sum_kw_k=1$ are known, and $M_\beta=\{\mu\in\mathbb{R}^K:\,0\le\mu_k\le w_k/\beta,\ \sum_k\mu_k=1\}$. For a vector $\tau\in\mathbb{R}^K$, order its entries as $\tau_{(1)}\le\cdots\le\tau_{(K)}$, let $w_{(i)}$ be the weight attached to $\tau_{(i)}$, let $W_i=\sum_{l\le i}w_{(l)}$ be the cumulative weights ($W_0=0$), and let $j_\beta=\min\{i:\,W_i\ge\beta\}$. Denote by $\mathbb{S}_\beta(\tau)$ and $\Lambda(\tau)$ the optimal value and the set of optimal $\lambda$ of program \eqref{LP_Superquantile} with $\hat\tau(X_k)$ replaced by $\tau_k$.
	
	\begin{lemma}\label{Lem:LP_KKT}
		The program \eqref{LP_Superquantile} has the dual $\min_{\mu\in M_\beta}\sum_k\mu_k\tau_k$, with the same value, and $(\lambda,z)$ is optimal if and only if there exist $\mu,\nu\in\mathbb{R}^K_+$ with $\sum_k\mu_k=1$, $\mu_k+\nu_k=w_k/\beta$, $\mu_k(\tau_k-\lambda-z_k)=0$ and $\nu_kz_k=0$ for all $k$, together with primal feasibility. Moreover,
		\[
		\mathbb{S}_\beta(\tau)=\frac{1}{\beta}\Big[\sum_{i=1}^{j_\beta-1}w_{(i)}\tau_{(i)}+(\beta-W_{j_\beta-1})\,\tau_{(j_\beta)}\Big],\qquad
		\Lambda(\tau)=\begin{cases}\{\tau_{(j_\beta)}\} & W_{j_\beta}>\beta,\\ [\tau_{(j_\beta)},\tau_{(j_\beta+1)}] & W_{j_\beta}=\beta.\end{cases}
		\]
		In particular, $\mathbb{S}_\beta(\cdot)$ is continuous, concave and piecewise linear, and it is differentiable at $\tau$ unless two coordinates with different tail status (weight $w_k/\beta$, an intermediate weight, or zero weight in the optimal $\mu$) coincide. With equal weights $w_k=1/K$, $j_\beta=\lceil\beta K\rceil$ and the interval case occurs exactly when $\beta K$ is an integer.
	\end{lemma}
	
	\proof
	The feasible set is nonempty and $M_\beta\neq\emptyset$, so strong duality holds and the stated conditions are the KKT system (Proposition~1 in \cite{Shum_2022}, specialized to the present constraint matrix). Stationarity in $\lambda$ and $z_k$ gives $\sum_k\mu_k=1$ and $\mu_k+\nu_k=w_k/\beta$. For the closed form, the objective in \eqref{S_estimates} is concave in $\lambda$ with slope $1-\beta^{-1}\sum_{k:\tau_k<\lambda}w_k$. The slope is positive for $\lambda\le\tau_{(j_\beta)}$, and on $(\tau_{(j_\beta)},\tau_{(j_\beta+1)})$ it equals $1-W_{j_\beta}/\beta$, which is negative if $W_{j_\beta}>\beta$ and zero if $W_{j_\beta}=\beta$.
	\eproof
	
	\subsubsection{Confidence sets.} Suppose $\tau(X_k,t)=\tau(X_k,t;\psi_P)$ for a finite-dimensional parameter $\psi_P\in\Xi\subset\mathbb{R}^p$, with $\Xi$ closed and $\tau(X_k,t;\cdot)$ continuous, and write $\mathbb{S}_\beta(\psi)$ and $\Lambda(\psi)$ for $\mathbb{S}_\beta$ and $\Lambda$ evaluated at $\tau(\psi)=(\tau(X_k,t;\psi))_{k\le K}$. The estimates in Section~\ref{Sec:Bounds_Ind_Welfare} are $\hat\tau(X_k)=\tau(X_k,t;\hat\psi)$. Throughout, the cells $X_1,\ldots,X_K$ are treated as fixed, so that $\mathbb{S}_\beta(\psi_P)$ is the superquantile of the $K$-point distribution that places weight $w_k$ on cell $k$; sampling error in the cells themselves is not accounted for.
	
	\begin{assumption}\label{Ass:psi_hat}
		Uniformly over $P\in\mathcal{P}$, $\sqrt{n}(\hat\psi-\psi_P)\xrightarrow{d}N(0,\Sigma_P)$, the eigenvalues of $\Sigma_P$ lie in $[\varepsilon,C]$ for some $0<\varepsilon<C<\infty$, and $\hat\Sigma_n\xrightarrow{p}\Sigma_P$ uniformly.
	\end{assumption}
	
	Let $C_n(1-\alpha)=\{\psi\in\Xi:\ n(\hat\psi-\psi)'\hat\Sigma_n^{-1}(\hat\psi-\psi)\le\chi^2_p(1-\alpha)\}$ and define
	\[
	CI_n(1-\alpha)=\Big[\min_{\psi\in C_n(1-\alpha)}\mathbb{S}_\beta(\psi),\ \max_{\psi\in C_n(1-\alpha)}\mathbb{S}_\beta(\psi)\Big],\qquad
	CS^\lambda_n(1-\alpha)=\bigcup_{\psi\in C_n(1-\alpha)}\Lambda(\psi).
	\]
	
	\begin{proposition}\label{Prop:LP_CI}
		Under Assumption~\ref{Ass:psi_hat},
		\[
		\liminf_{n\to\infty}\ \inf_{P\in\mathcal{P}}\ \Pr\nolimits_P\Big(\mathbb{S}_\beta(\psi_P)\in CI_n(1-\alpha)\ \text{and}\ \Lambda(\psi_P)\subseteq CS^\lambda_n(1-\alpha)\Big)\ \ge\ 1-\alpha .
		\]
		The coverage guarantee holds whether or not the cumulative weight $W_{j_\beta}$ equals $\beta$ and irrespective of ties among the $\tau(X_k,t;\psi_P)$.
	\end{proposition}
	
	\proof
	By Assumption~\ref{Ass:psi_hat}, $\Pr_P(\psi_P\in C_n(1-\alpha))\to1-\alpha$ uniformly in $P$. On the event $\{\psi_P\in C_n(1-\alpha)\}$, $\mathbb{S}_\beta(\psi_P)$ lies between the minimum and the maximum of $\mathbb{S}_\beta(\cdot)$ over $C_n(1-\alpha)$, and $\Lambda(\psi_P)\subseteq\bigcup_{\psi\in C_n}\Lambda(\psi)$. The extrema are attained because $C_n(1-\alpha)$ is compact ($\Xi$ is closed) and $\mathbb{S}_\beta(\cdot)$ is continuous by Lemma~\ref{Lem:LP_KKT}.\eproof
	
	\begin{remark}
		Proposition~\ref{Prop:LP_CI} uses neither the differentiability of $\mathbb{S}_\beta$ nor a rank condition on the joint covariance of $(\hat\tau(X_k))_{k\le K}$, which is singular whenever $p<K$; this is why it is stated at the level of $\hat\psi$. The price is conservativeness, since a $p$-dimensional confidence set is projected onto a scalar. Three further points are worth recording.
		\begin{enumerate}
			\item[(i)] \emph{Scope.} Assumption~\ref{Ass:psi_hat} is high-level: it covers parametric and other $\sqrt{n}$-estimable specifications of $\tau(X,t)$, but not estimates of $\tau(X,t)$ obtained by nonparametric methods at slower rates.
			\item[(ii)] \emph{Computation.} Validity requires the exact extrema of $\mathbb{S}_\beta(\cdot)$ over $C_n(1-\alpha)$, so a local solver, which can return an interval that is too short, is not adequate. Since $\mathbb{S}_\beta(\psi)=\min_{\mu\in M_\beta}\sum_k\mu_k\tau(X_k,t;\psi)$, the maximum over $C_n(1-\alpha)$ is a convex problem when $\tau(X_k,t;\cdot)$ is linear, whereas the minimum is a concave minimization; for small $K$ it can be computed exactly by enumerating the vertices of $M_\beta$, or with a global solver.
			\item[(iii)] \emph{One-sided statement.} Applying Theorem~\ref{Bound_Change_Utility} to the $K$-point distribution above gives $\mathbb{S}_\beta(\tau(X,t,\varepsilon))\le\mathbb{S}_\beta(\psi_P)$, so the upper endpoint of $CI_n(1-\alpha)$ is an upper confidence limit for the superquantile of $\tau(X,t,\varepsilon)$ itself with asymptotic coverage at least $1-\alpha$.
		\end{enumerate}
		If instead the $\hat\tau(X_k)$ are estimated separately with a nonsingular joint covariance $\Omega_P$, an alternative set follows the primal--dual construction of \cite{Shum_2022}, with the deterministic constraints $z\le0$ and $\mu\in M_\beta$ imposed as constraints rather than moments; we do not pursue it here.
	\end{remark}

	
	\section{Proofs}\label{Sec:Proofs_WDZ}
	\subsection{Proof of Lemma \ref{phi_dual_lemma}}
	This is the dual of a convex optimization problem over $G$ subject to the linear constraint $\EE_F[1]=1$ and the convex constraint $D_\phi(G\|F)\leq \rho$; strong duality holds under the stated conditions on $\phi$ by Proposition 1 and the duality theorem in \cite{BentalTeboulle1987}. \eproof
	
	\subsection {Proof of Proposition \ref{Robust_CVAR_Dual_Prop}}
	Immediate from Lemma \ref{phi_dual_lemma} applied pointwise in $\lambda$, since $(Z-\lambda)_-$ is bounded below (by $-|\lambda|-\EE_F|Z|$, say, under the integrability maintained throughout) and concavity in $(\lambda,\eta,\kappa)$ follows because $\phi^*$ is convex and the perspective of a convex function is jointly convex in $(\eta-(Z-\lambda)_-,\kappa)$.\eproof
	
	\subsection{Proof of Theorem \ref{Bound_Change_Utility}}
	By definition, we know that $$\mathbb{S}_\beta(\tau(X,t,\varepsilon))=\sup_{\lambda}\left\{\lambda+{1\over \beta}\EE[\min\{\tau(X,t,\varepsilon)-\lambda,0\}]\right\}.$$
	By the law of iterated expectations, the previous expression can be rewritten  as:
	$$\mathbb{S}_\beta(\tau(X,t,\varepsilon))=\sup_{\lambda}\left\{\lambda+{1\over \beta}\EE[\EE[\min\{\tau(X,t,\varepsilon)-\lambda,0\}|X]]\right \}.$$
	By applying Jensen's inequality (to the concave function $\min\{\}$), we get:
	$$\mathbb{S}_\beta(\tau(X,t,\varepsilon))\leqslant \sup_{\lambda}\left\{\lambda+{1\over \beta}\EE[\min\{\tau(X,t)-\lambda,0\}|X]\right \}.$$
	Applying the  definition of superquantile to the random variable $\tau(X,t)$, we conclude that $\mathbb{S}_{\beta}(\tau(X,t,\varepsilon))\leqslant \mathbb{S}_{\beta}(\tau(X,t))$ for all $\beta\in(0,1]$.
	
	\eproof
	
	\subsection{Proof of Corollary \ref{Quant_Aggregate_Bound_Change_W}}
	The result follows from  the fact that (\cite[Thm. 3.1]{Royset2025}) $$\mathbb{S}_\beta(\tau(X,t,\varepsilon))={1\over \beta}\int_{0}^\beta F^{-1}_{\tau(X,t,\varepsilon)}(\theta)d \theta$$
	and
	$$\mathbb{S}_\beta(\tau(X,t))={1\over \beta}\int_{0}^\beta F^{-1}_{\tau(X,t)}(\theta)d \theta.$$
	Putting these two expressions together we get the conclusion in the corollary.\eproof
	
	\subsection{Proof of Corollary \ref{Cor:SSD}}
	By Theorem \ref{Bound_Change_Utility}, $\mathbb{S}_\beta(\tau(X,t,\varepsilon))\leq \mathbb{S}_\beta(\tau(X,t))$ for every $\beta\in(0,1]$. Multiplying both sides by $\beta>0$ and applying the identity $\beta\,\mathbb{S}_\beta(Z)=\int_{0}^\beta F^{-1}_{Z}(\theta)d\theta$ (\cite[Thm. 3.1]{Royset2025}) to $Z=\tau(X,t,\varepsilon)$ and $Z=\tau(X,t)$ in turn, we obtain
	$$\int_{0}^\beta F^{-1}_{\tau(X,t,\varepsilon)}(\theta)\,d\theta \;\leq\; \int_{0}^\beta F^{-1}_{\tau(X,t)}(\theta)\,d\theta \qquad\text{for every }\beta\in(0,1].$$
	This is precisely the condition for $\tau(X,t)$ to second-order stochastically dominate $\tau(X,t,\varepsilon)$, which is equivalent to $\EE[u(\tau(X,t,\varepsilon))]\leq \EE[u(\tau(X,t))]$ for every nondecreasing, concave $u$ (\cite[Thm. 3.1]{OgryczakRuszczynski2002}).\eproof
	
	\subsection{Proof of Proposition \ref{Prop:bPOE}}
	Write $f(\beta)=\mathbb{S}_\beta(\tau(X,t,\varepsilon))$ and $g(\beta)=\mathbb{S}_\beta(\tau(X,t))$. By Theorem \ref{Bound_Change_Utility}, $f(\beta)\leq g(\beta)$ for every $\beta\in(0,1]$. If $f(\beta)\geq \theta$ for some $\beta$, then $g(\beta)\geq f(\beta)\geq \theta$ as well; hence $\{\beta:f(\beta)\geq \theta\}\subseteq\{\beta:g(\beta)\geq \theta\}$, and taking the infimum of each set reverses the inclusion:
	$$p_\theta(\tau(X,t))=\inf\{\beta:g(\beta)\geq \theta\}\;\leq\;\inf\{\beta:f(\beta)\geq \theta\}=p_\theta(\tau(X,t,\varepsilon)).$$
	\eproof
	
	\subsection{Proof of Theorem \ref{Bound_Change_Utility_Lower}}  We note that $\tau(X,t)-\tau(X,t,\varepsilon)\leq \gamma$ can be rewritten as 
	$\tau(X,t,\varepsilon)\geq \tau(X,t)-\gamma$. Then, applying the definition of superquantile, we get:
	\begin{eqnarray*}
		\mathbb{S}_\beta(\tau(X,t,\varepsilon))&=&\sup_{\lambda}\left\{\lambda+{1\over \beta}\EE[\min(\tau(X,t,\varepsilon)-\lambda)_-]\right \}\\
		&\geq &\sup_{\lambda}\left\{\lambda+{1\over \beta}\EE[(\tau(X,t))-\gamma-\lambda)_{-}]\right\}\\
		&=&\sup_{\tilde{\lambda}}\left\{(\lambda+\gamma)+{1\over \beta}\EE[(\tau(X,\varepsilon))-(\lambda+\gamma))_{-}]\right\}-\gamma\\
		&=&\mathbb{S}_\beta(\tau(X,t))-\gamma,
	\end{eqnarray*}
	where the last equality used the fact $\tilde{\lambda}=\lambda+\mu$ and $(X)_-\triangleq\min\{X,0\}$
	
	\eproof
	
	\subsection{Proof of Theorem \ref{Robust_Bound_Change_Utility}}
	Fix $G\in\mathcal{U}_\phi(F,\rho)$. By the law of iterated expectations (under $G\otimes F_{\varepsilon\mid X}$) and Jensen's inequality applied to the concave map $z\mapsto[z-\lambda]_-$ --- exactly as in the proof of Theorem \ref{Bound_Change_Utility} in Appendix \ref{Sec:Proofs_WDZ}, and using only the conditional law $F_{\varepsilon\mid X}$, which is unaffected by the choice of $G$ --- we obtain, for every $\lambda\in\RR$,
	$$\EE_{F_{\varepsilon\mid X}}\big[(\tau(X,t,\varepsilon)-\lambda)_-\mid X\big]\leq (\tau(X,t)-\lambda)_-.$$
	Taking $\EE_G[\cdot]$ on both sides (a positive linear operator, so the inequality is preserved) and adding $\lambda$,
	$$\lambda+\frac1\beta\,\EE_G\Big[\EE_{F_{\varepsilon\mid X}}\big[(\tau(X,t,\varepsilon)-\lambda)_-\mid X\big]\Big]\leq \lambda+\frac1\beta\,\EE_G\big[(\tau(X,t)-\lambda)_-\big].$$
	Taking $\sup_{\lambda\in\RR}$ of both sides preserves the inequality, giving $\mathbb{S}_\beta^G(\tau(X,t,\varepsilon))\leq \mathbb{S}_\beta^G(\tau(X,t))$. Since $G\in\mathcal{U}_\phi(F,\rho)$ was arbitrary, taking $\inf_{G\in\mathcal{U}_\phi(F,\rho)}$ of both sides yields (\ref{Robust_Aggregate_Bound_Change_W}).\eproof
	
	\subsection{Proof of Corollary \ref{Cor:bPOE_Robust}}
	The argument is identical to the proof of Proposition \ref{Prop:bPOE} in Section \ref{Sec:Bounds_Ind_Welfare} above, with $\mathbb{S}_\beta(\cdot)$ replaced throughout by $\mathbb{S}_\beta^R(\cdot)$ and Theorem \ref{Bound_Change_Utility} replaced by Theorem \ref{Robust_Bound_Change_Utility}: writing $f(\beta)=\mathbb{S}_\beta^R(\tau(X,t,\varepsilon))$ and $g(\beta)=\mathbb{S}_\beta^R(\tau(X,t))$, Theorem \ref{Robust_Bound_Change_Utility} gives $f(\beta)\leq g(\beta)$ for every $\beta\in(0,1]$, so $\{\beta:f(\beta)\geq \theta\}\subseteq\{\beta:g(\beta)\geq \theta\}$ and taking infima yields $p_\theta^R(\tau(X,t);\rho)\leq p_\theta^R(\tau(X,t,\varepsilon);\rho)$.\eproof
	
	\subsection{Proof of Theorem \ref{Robust_Bound_Change_Utility_Lower}.}
	Fix $G\in\mathcal{U}_\phi(F,\rho)$. The hypothesis $\tau(X,t)-\tau(X,t,\varepsilon)\leq\gamma$ is a restriction on $F_{\varepsilon\mid X}$ and on the fixed function $\tau(\cdot,t)$ only; it does not involve the marginal law of $X$, and therefore continues to hold a.s.\ under $G\otimes F_{\varepsilon\mid X}$ for every $G$. The argument in the proof of Theorem \ref{Bound_Change_Utility_Lower} (Appendix \ref{Sec:Proofs_WDZ}) therefore applies verbatim under $G\otimes F_{\varepsilon\mid X}$ and yields
	$$\mathbb{S}_\beta^G(\tau(X,t,\varepsilon))\geq \mathbb{S}_\beta^G(\tau(X,t))-\gamma\qquad\text{for every }G\in\mathcal{U}_\phi(F,\rho).$$
	Taking $\inf_{G\in\mathcal{U}_\phi(F,\rho)}$ of both sides and using that $\gamma$ does not depend on $G$ (so the infimum of the right-hand side equals $\mathbb{S}_\beta^R(\tau(X,t))-\gamma$ exactly, with no slack), we obtain (\ref{Robust_Aggregate_Bound_Change_W_Lower}).\eproof
	
	\subsection{Proof of Proposition \ref{Bound_Welfare_MTE}}  Bound (\ref{Gen_Roy_B1}) is a direct consequence of Theorem \ref{Bound_Change_Utility}.  To establish (\ref{Gen_Roy_B2}), note that by Assumption \ref{Generalized_Roy_Assumpton}(iii) the distribution function $F_{\varepsilon_{\mathcal{W}}}$ is continuous and strictly increasing, so that $\varepsilon_{\mathcal{W}}=F^{-1}_{\varepsilon_{\mathcal{W}}}(U_{\mathcal{W}})$ almost surely. Therefore
	$$\mathcal{W}=\mu_{\mathcal{W}}(X,Z)-F^{-1}_{\varepsilon_{\mathcal{W}}}(U_{\mathcal{W}})$$
	is $\sigma(X,Z,U_{\mathcal{W}})$-measurable, and $\mathcal{W}^{\operatorname{MTE}}(X,Z,U_{\mathcal{W}})=\EE[\mathcal{W}\mid X,Z,U_{\mathcal{W}}]=\mathcal{W}$ almost surely. Since $\mathbb{S}_\beta$ depends only on the distribution of its argument, $\mathbb{S}_{\beta}(\mathcal{W})=\mathbb{S}_\beta(\mathcal{W}^{\operatorname{MTE}}(X,Z,U_{\mathcal{W}}))$. 
	\eproof
	
	\subsection{Proof of Proposition \ref{C_lower_bound}}The proofs of bounds (\ref{Lower_bound_cost_ATE}) and (\ref{Lower_bound_cost_MTE}) are identical. For ease of exposition, we present only the proof of bound (\ref{Lower_bound_cost_ATE}).
	\smallskip
	
	By definition we know that
	$$\overline{\mathbb{S}}_\alpha(C)=\sup_{\lambda}\left\{\lambda+{1\over 1-\alpha} \EE[(C-\lambda)_+]\right\}.$$ By the law of iterated expectation we have:
	$$\overline{\mathbb{S}}_\alpha(C)=\sup_{\lambda}\left\{\lambda+{1\over 1-\alpha} \EE[\EE[(C-\lambda)_+\mid Z]]\right\}$$.
	By Jensen's inequality (convex case) and using the definition of $\mu_C(Z)$ we get:
	$$\overline{\mathbb{S}}_\alpha(C)\geq \sup_{\lambda}\left\{\lambda+{1\over 1-\alpha} \EE[(\mu_C(Z)-\lambda)_+]\right\}.$$
	By the definition of superquantile, we conclude that:
	$$\overline{\mathbb{S}}_\alpha(C)\geq\overline{\mathbb{S}}_\alpha(C^{\operatorname{ATE}}(Z))\quad \text{for all $\alpha\in[0,1)$}.$$
	\eproof
	
	\subsection{Proof of Theorem \ref{Wel_Cost_TT_bounds}} 
	(i) By definition, we know that
	\begin{eqnarray*}
		\mathbb{S}_\beta(\mathcal{W}^{\operatorname{TT}})&=&\sup_{\lambda \in \RR}\left\{\lambda+{1\over\beta}\EE[(\mathcal{W}-\lambda)_{-}\mid D=1]\right\}.
	\end{eqnarray*}
	By the law of iterated expectation, we get 
	\begin{eqnarray*}
		\mathbb{S}_\beta(\mathcal{W}^{\operatorname{TT}})&=&\sup_{\lambda \in \RR}\left\{\lambda+{1\over\beta}\EE[\EE[(\mathcal{W}-\lambda)_{-}\mid X,Z, D=1]]\right\}.
	\end{eqnarray*}
	
	Using Jensen's inequality, we get 
	\begin{eqnarray*}
		\mathbb{S}_\beta(\mathcal{W}^{\operatorname{TT}})&\leq &\sup_{\lambda \in \RR}\left\{\lambda+{1\over\beta}\EE[\EE[\mathcal{W}\mid X,Z, D=1]-\lambda)_{-}]\right\}.
	\end{eqnarray*}
	Using the definition of $\mathcal{W}^{\operatorname{TT}}(X,Z)$ we conclude 
	$$\mathbb{S}_\beta(\mathcal{W}^{TT})\leq \mathbb{S}_\beta(\mathcal{W}^{\operatorname{TT}}(X,Z))\quad \text{for all $\beta\in(0,1].$}$$

	(ii) By the definition of $\overline{\mathbb{S}}_\alpha(C^{\operatorname{TT}})$ we know
	\begin{eqnarray*}
		\overline{\mathbb{S}}_\alpha(C^{\operatorname{TT}})&=&\inf_{\lambda\in\RR}\left\{\lambda+{1\over 1-\alpha}\EE[(C-\lambda)_{+}\mid D=1]\right\}.
	\end{eqnarray*}
	Using the law of iterated expectation and Jensen's inequality once again, we find that:
	\begin{eqnarray*}
		\overline{\mathbb{S}}_\alpha(C^{\operatorname{TT}})&\geq &\inf_{\lambda\in\RR}\left\{\lambda+{1\over 1-\alpha}\EE[(C\mid Z,D=1)-\lambda)_{+}]\right\}.
	\end{eqnarray*}
	Combining the previous inequality with the definition of the superquantile we obtain:
	$$\overline{\mathbb{S}}_\alpha(C^{\operatorname{TT}})\geq \overline{\mathbb{S}}_\alpha(C^{\operatorname{TT}}(Z)) \quad\text{for all $\alpha\in[0,1).$}$$
	\eproof
	
	\subsection{Proof of Corollary \ref{Cor:bPOE_TT}}
	The argument is identical to the proof of Proposition \ref{Prop:bPOE}, with $\tau(X,t,\varepsilon)$ and $\tau(X,t)$ replaced by $\mathcal{W}^{TT}$ and $\mathcal{W}^{TT}(X,Z)$, and Theorem \ref{Bound_Change_Utility} replaced by Theorem \ref{Wel_Cost_TT_bounds}(i): writing $f(\beta)=\mathbb{S}_\beta(\mathcal{W}^{TT})$ and $g(\beta)=\mathbb{S}_\beta(\mathcal{W}^{TT}(X,Z))$, Theorem \ref{Wel_Cost_TT_bounds}(i) gives $f(\beta)\leq g(\beta)$ for every $\beta\in(0,1]$, so $\{\beta:f(\beta)\geq \theta\}\subseteq\{\beta:g(\beta)\geq \theta\}$ and taking infima yields $p_\theta(\mathcal{W}^{TT}(X,Z))\leq p_\theta(\mathcal{W}^{TT})$.\eproof
	\newpage
\end{document}